\documentclass[superscriptaddress,twocolumn,aps,prl, reprint,showpacs]{revtex4-1}

\usepackage{epsfig,graphicx,times}
\usepackage[utf8]{inputenc}

\usepackage{pgf}
\usepackage{braket}
\usepackage{xr}
\usepackage{pdfpages}
\graphicspath{{figures_7_6_16/}}

\DeclareUnicodeCharacter{00A0}{ }

\begin{document}
\title{Observation of tunneling-assisted highly forbidden single-photon
transitions in a Ni$_4$ single-molecule magnet}

\author{Yiming Chen}
\affiliation{Department of Physics and Astronomy, Amherst College, Amherst, MA 01002-5000, USA}
\affiliation{Department of Physics, University of Massachusetts, Amherst, MA 01003, USA}
\author{Mohammad D. Ashkezari}
\altaffiliation[Current address: ]{Department of Earth, Atmospheric and Planetary Sciences, Massachusetts Institute of Technology, Cambridge, Massachusetts 02139, USA}
\affiliation{Department of Physics and Astronomy, Amherst College,
Amherst, MA 01002-5000, USA}
\author{Charles A. Collett}
\affiliation{Department of Physics and Astronomy, Amherst College, Amherst, MA 01002-5000, USA}

\author{Rafael A. All\~{a}o Cassaro}
\altaffiliation[Current address: ]{Instituto de Qu\'{\i}mica, Universidade Federal do Rio de Janeiro, Rio de Janeiro, RJ, 21945-970  Brazil}
\affiliation{Department of Chemistry, University of Massachusetts, Amherst, MA 01003, USA}

\author{Filippo Troiani}
\affiliation{S3 Instituto Nanoscienze-CNR, I-41124 Modena, Italy}

\author{Paul M. Lahti}
\affiliation{Department of Chemistry, University of Massachusetts,
Amherst, MA 01003, USA}

\author{Jonathan R. Friedman}
\email[Corresponding author: ]{jrfriedman@amherst.edu}
\affiliation{Department of Physics and Astronomy, Amherst College, Amherst, MA 01002-5000, USA}

\date{\today}

\begin{abstract}
Forbidden transitions between energy levels typically involve violation of selection rules imposed by symmetry and/or conservation laws.  A nanomagnet tunneling between up and down states violates angular momentum conservation because of broken  rotational symmetry.  Here we report observations of highly forbidden transitions between spin states in a Ni$_4$ single-molecule magnet in which a single photon can induce the spin to change by several times $\hbar$, nearly reversing the direction of the spin.  These observations are understood as tunneling-assisted transitions that lift the standard $\Delta m = \pm 1$ selection rule for single-photon transitions.  These transitions are observed at low applied fields, where tunneling is dominated by the molecule's intrinsic anisotropy and the field acts as a perturbation.  Such transitions can be exploited to create macroscopic superposition states that are not typically accessible through single-photon $\Delta m = \pm 1$ transitions.
\end{abstract}

\maketitle

There has been much recent attention to using spin systems as potential qubits~\cite{schuster_high-cooperativity_2010,kubo_strong_2010,wolfowicz_atomic_2013,bader_room_2014}. Molecular nanomagnets  are particularly attractive as spin qubits~\cite{leuenberger_quantum_2001,ardavan_will_2007,wedge_chemical_2012,schlegel_direct_2008,bogani_molecular_2008,troiani_molecular_2011,ghosh_multi-frequency_2012,bader_room_2014,shiddiq_enhancing_2016} because many of their properties can be chemically engineered. Single-molecule magnets (SMMs) are anisotropic molecular magnets, typically with large total spin, for which the spin is impelled to point along a preferred axis, the ``easy" axis~\cite{friedman_single-molecule_2010}. They exhibit remarkable quantum dynamics including tunneling between different orientations~\cite{friedman_macroscopic_1996} and quantum-phase interference~\cite{wernsdorfer_quantum_1999}.
Here we present evidence of highly forbidden transitions in the Ni$_4$ SMM where the transitions are enabled by tunneling, which lifts the requirement of spin angular momentum conservation.  We observe transitions in which the absorption of a single photon permits a near reversal of the molecule's macrospin, grossly violating the standard $\Delta m = \pm 1$ selection rule.  The quantum states that can be generated through these forbidden transitions are non-classical, having a substantial ``macroscopicity" by a standard measure.  Our results imply that the forbidden transitions observed in this system (and similar molecules with strong anisotropy) can be exploited to create highly nonclassical states with single-photon transitions.

From a quantum coherence perspective, forbidden transitions have some distinct advantages: Since the matrix elements for these transitions are small, they tend to have long lifetimes.  In addition, they can be less susceptible to magnetic-field fluctuations under certain circumstances, potentially leading to longer coherence times~\cite{bollinger_laser-cooled-atomic_1985,wolfowicz_atomic_2013,shiddiq_enhancing_2016}.
Forbidden transitions have been seen in SMMs
with very strong tunneling produced by strongly broken symmetry~\cite{ghosh_multi-frequency_2012,delbarco_quantum_2004,shiddiq_enhancing_2016}. In contrast, in our experiments the transitions are dominated by a modest intrinsic anisotropy with an applied field acting as a perturbation.

\begin{figure}[h]
\centering
\includegraphics[width=1\linewidth]{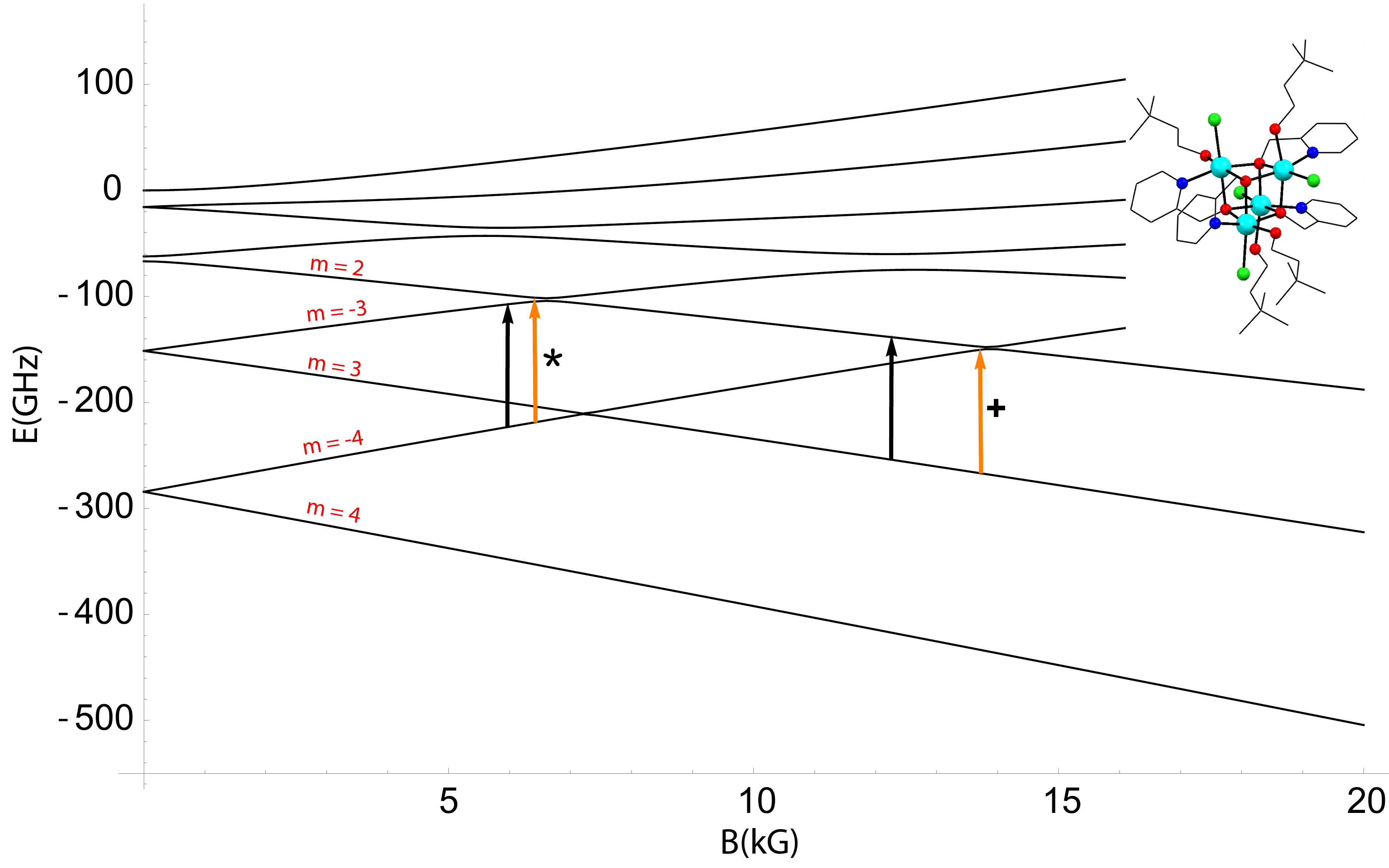}
\caption{Spin-state energy-level diagram for one conformational state (``black" -- see below) of Ni$_4$.  Energies of various levels are shown as a function of magnetic field, calculated by diagonalizing the molecule's spin Hamiltonian.  The diagram illustrates the levels' behavior when $\theta = 30^\circ$.  Arrows indicate the major transitions observed in this study: Black = allowed, orange = forbidden. The two orange arrows are labeled with $\star$ and $+$, the designations used throughout this article. Inset:  Molecular unit of [Ni(hmp)(dmb)Cl]$_4$~\cite{yang_exchange_2003}, where hmp is the anion of 2-hydroxymethylpyridine  and dmb is 3,3-dimethyl-1-butanol. Color code: green--chloride; cyan--nickel(II); black--carbon; red--oxygen; blue--nitrogen. Hydrogens have been omitted for clarity.}
\label{Energygraph}
\end{figure}

We studied the $S=4$ complex [Ni(hmp)(dmb)Cl]$_4$ (hereafter Ni$_4$), shown in the inset of Fig.~\ref{Energygraph}.  The molecule's large ligands  isolate the magnetic centers within a crystal from each other~\cite{yang_exchange_2003}. In addition, there are no solvate molecules in the crystal lattice and 99\% (natural abundance) of Ni nuclei have spin $I=0$. This SMM has been characterized by electron-spin resonance (ESR) spectroscopy~\cite{edwards_high-frequency_2003,delbarco_quantum_2004,kirman_origin_2005,de_loubens_magnetization_2007,de_loubens_high_2008,lawrence_disorder_2008,lawrence_magnetic_2009},
magnetization measurements~\cite{yang_exchange_2003,delbarco_quantum_2004,yang_fast_2006}
and heat capacity measurements~\cite{hendrickson_origin_2005,lawrence_disorder_2008,beedle_ferromagnetic_2010}.
Ni$_4$  can be well described as a single ``giant spin" with the Hamiltonian~\cite{lawrence_magnetic_2009}:
\begin{equation}
H = -DS_z^2 - AS_z^4 + C(S_+^4 + S_-^4 ) - \mu _B \vec B \cdot
\mathbf{g} \cdot \vec S,
\label{Ni4Ham}
\end{equation}
\noindent
where $\mathbf{g}$ is the molecule's g tensor, $D$ and $A$ are axial (diagonal) anisotropy parameters that define the ``easy" z axis and make the \hbox{$m=\pm 4$} magnetic sublevels have the lowest energy, producing an energy barrier between those two orientations; $C$ is a transverse (off-diagonal) anisotropy parameter that affects the strength of tunneling through the barrier; and the magnetic field  $\vec B=B\left(\sin\theta\cos\phi,\sin\theta\sin\phi,\cos\theta\right)$ produces a Zeeman interaction. The z component of $\vec B$  changes the energies of the magnetic sublevels as illustrated in Fig.~\ref{Energygraph}. When levels approach, the off-diagonal terms in Eq.~\ref{Ni4Ham} mix states of different $m$ values, giving rise to anticrossings. Like the transverse anisotropy, the transverse components of $\vec B$ are off-diagonal terms in Eq.~\ref{Ni4Ham}.  Since the off-diagonal terms do not commute with $S_z$, they are responsible for the observed tunneling phenomena in this and other SMMs~\cite{friedman_macroscopic_1996,friedman_single-molecule_2010}.  The energy splitting at an anticrossing is dubbed the ``tunnel splitting".

We performed reflection ESR spectroscopy using a 3D cylindrical resonant cavity  with a TE$_{011}$ mode  with resonant frequency $\sim$115.54 GHz and a quality factor ($Q$) of $\sim$10000.  A static magnetic field $\vec{H}$ was applied
along the axis of the cavity. A single crystal of Ni$_4$ (synthesized using published procedures~\cite{yang_fast_2006}) was mounted on the bottom of
the cavity at a position where the rf field was perpendicular to the static field. The easy axis
of the crystal was manually tilted at various angles ($\theta_H$) relative to $\vec{H}$. We measured the reflected power as a function of frequency and extracted the resonance frequency and $Q$ value of the cavity at each field~\cite{supp}.

\begin{figure}[h]
\centering
\includegraphics[width=1\linewidth]{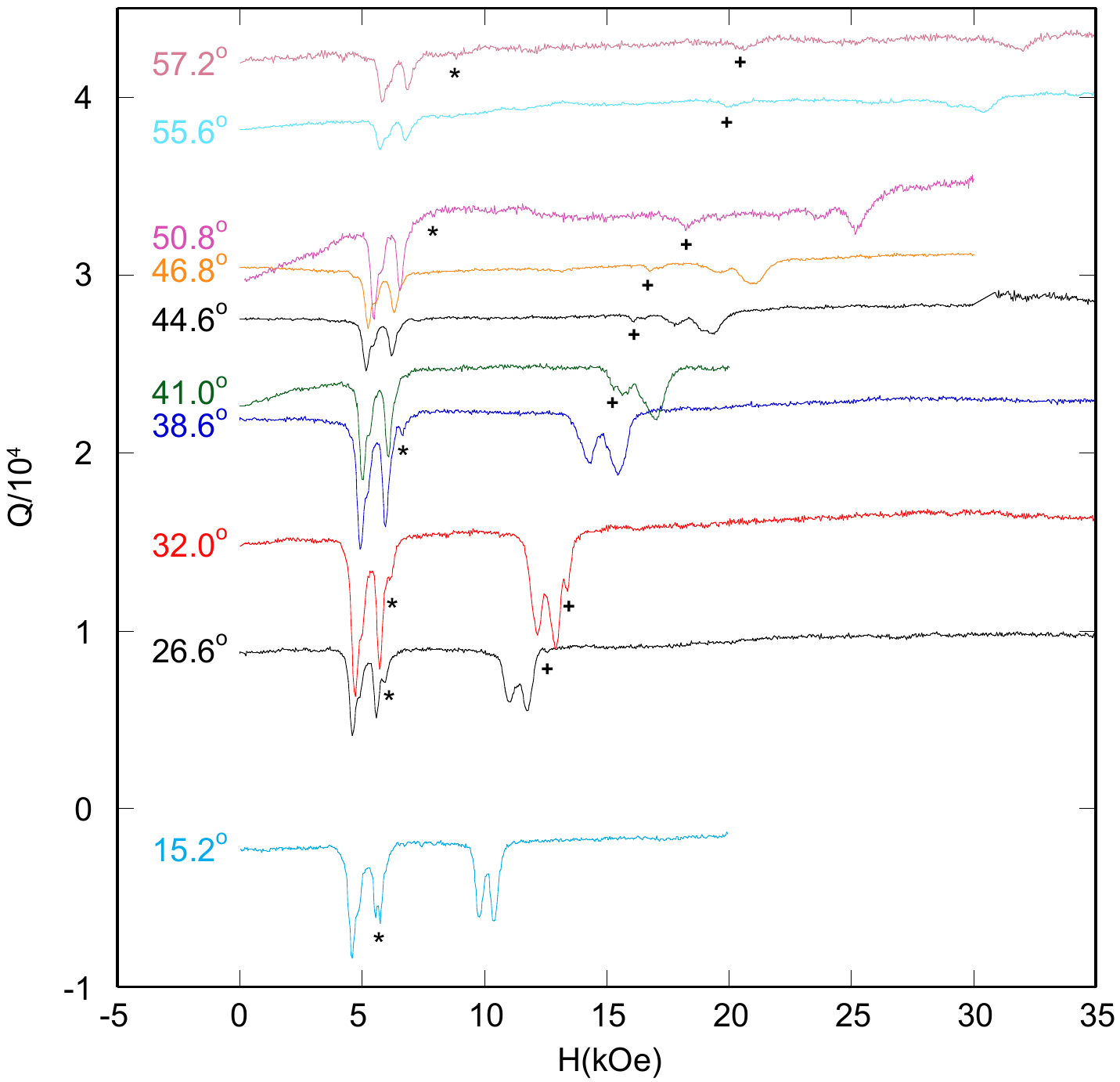}
\caption{Absorption ESR spectra at 1.8 K for several angles $\theta_H$.
The spectrum for $\theta_{H,ref}=26.6^\circ$ shows actual $Q$ values.  All other spectra have been shifted
vertically by an amount proportional to $\theta_H-\theta_{H,ref}$.  Spectra from three different crystals are combined in this figure.  Each spectrum has been shifted slightly horizontally to account for inductive effects due to sweeping $H$ (see~\cite{supp}).}
\label{spectrum}
\end{figure}

Figure \ref{spectrum} shows ESR spectra ($Q$ vs.~$H$) at 1.8 K for a single crystal of Ni$_4$ at multiple values of $\theta_H$, the angle between the easy axis and $\vec{H}$.
We typically observe multiple peaks: two large peaks that are each split and, often, small peaks to the right or left of the large peaks. Dispersive spectra show corresponding features (see Fig.~2
in~\cite{supp}). The large peaks correspond to allowed transitions with $\Delta m \simeq\pm1$.  The splitting of these peaks arises from ligand conformational disorder~\cite{hendrickson_origin_2005}. Additional fine structure that some of these peaks exhibit~\cite{lawrence_disorder_2008} is not relevant to this study. We focus on the two small side peaks (marked $\star$ and $+$ in Fig.~\ref{spectrum}) that correspond roughly to $m=-4 \to m= 2 $ ($\star$) and $m = 3 \to m = -4$ ($+$) (cf.~Fig.~\ref{Energygraph}, orange arrows).  Compared with the allowed transitions, these forbidden transitions have markedly different dependences on $\theta_H$, confirming their different character.

\begin{figure}[h]
\centering
\includegraphics[width=1\linewidth]{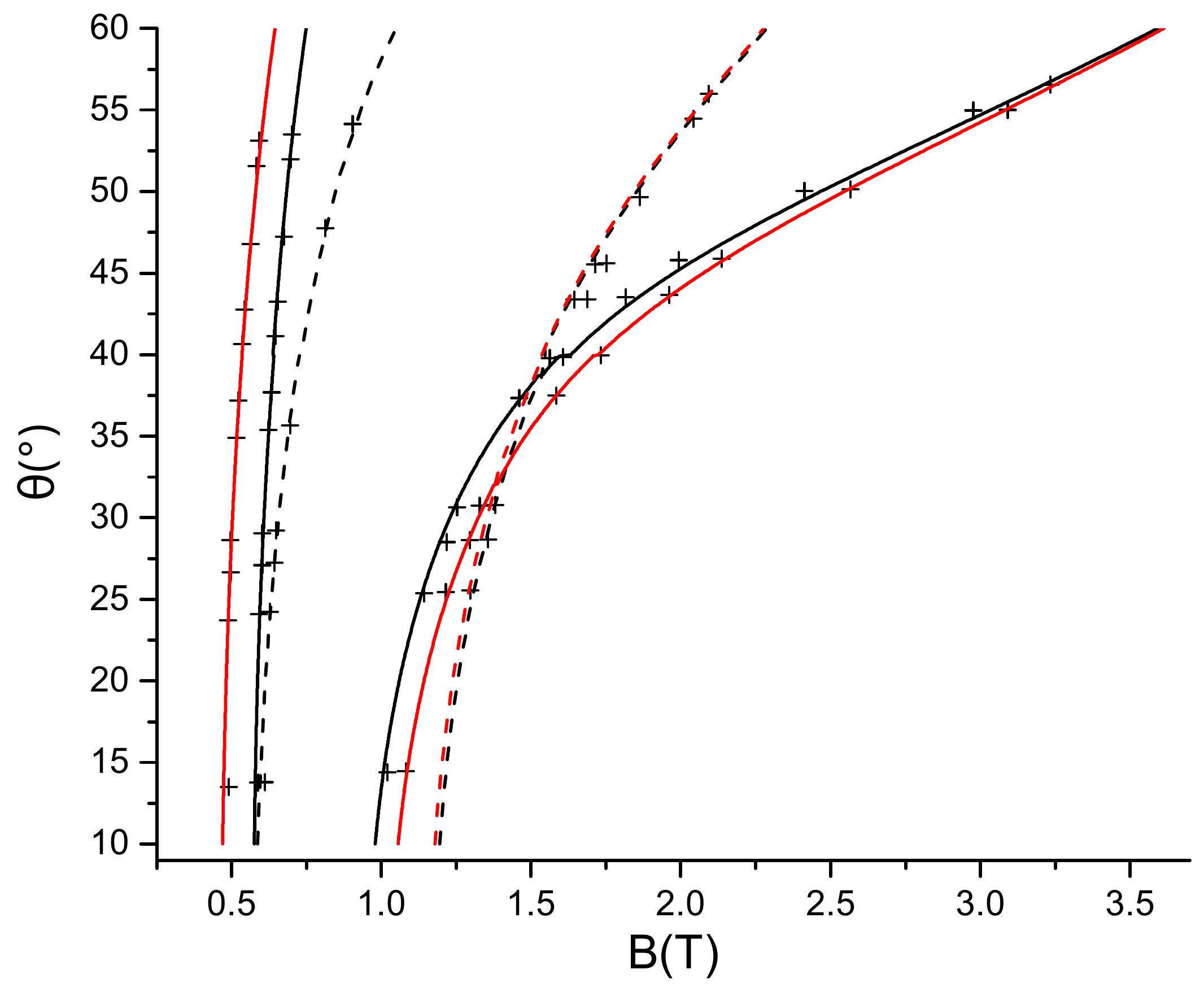}
\caption{Resonance positions in $B-\theta$ space.  The points are the peak positions from Fig.~\ref{spectrum} after correcting for the effects of dipole fields~\cite{supp}.  The lines are the results of simulations after fitting the observed spectra.  Black and red correspond to different conformational states of the molecule with correspondingly different anisotropy constants.  Solid curves indicate allowed transitions and dashed curves correspond to forbidden transitions.  The small shift seen in the calculated results at $\sim40^\circ$ arises from use of different samples at angles above and below this value and associated differences in the direction ($\phi$) of the transverse field in the samples' hard planes~\cite{supp}.}
\label{BvsTheta}
\end{figure}

Figure \ref{BvsTheta} shows the $B-\theta$ resonance positions (determined from the spectra in Fig.~\ref{spectrum}), where $\theta$ is the angle between the easy axis and the field $\vec{B}$ experienced by the molecules.  Lines show the calculated resonance points for the transitions shown in Fig.~\ref{Energygraph}, obtained by diagonalizing Eq.~\ref{Ni4Ham} using the parameters given below. Solid (dashed) curves indicate allowed (forbidden) transitions. The agreement between the calculated $B-\theta$ resonance positions and the experimental data is very good.  In producing Fig.~\ref{BvsTheta}, we took into account that both the magnitude and direction of $\vec{B}$ changes with $\vec{H}$ due to intermolecular dipolar interactions, so that each spectrum in Fig.~\ref{spectrum} produces a range of $\theta$ values in Fig.~\ref{BvsTheta}~\cite{supp}.  Red and black curves  show predicted resonance positions for the two conformational states (isomers) of the molecule, which have somewhat different anisotropy constants, determined by fitting~\cite{supp}:   $D=15.13(4)$ GHz, $A=0.136(2)$ GHz and $C=5.3(2)$ MHz (red), and $D=15.55(4)$ GHz, $A=0.138(2)$ MHz, $C=6.45(3)$ MHz (black). $g$ factors were taken to be the same for both components  and found to be $g_z=$ 2.157(7) and $g_x=g_y=$ 2.220(3).   These numbers are in reasonable agreement with those found by others~\cite{edwards_high-frequency_2003,kirman_origin_2005}.

\begin{figure}[!htb]
\includegraphics[width=1\linewidth]{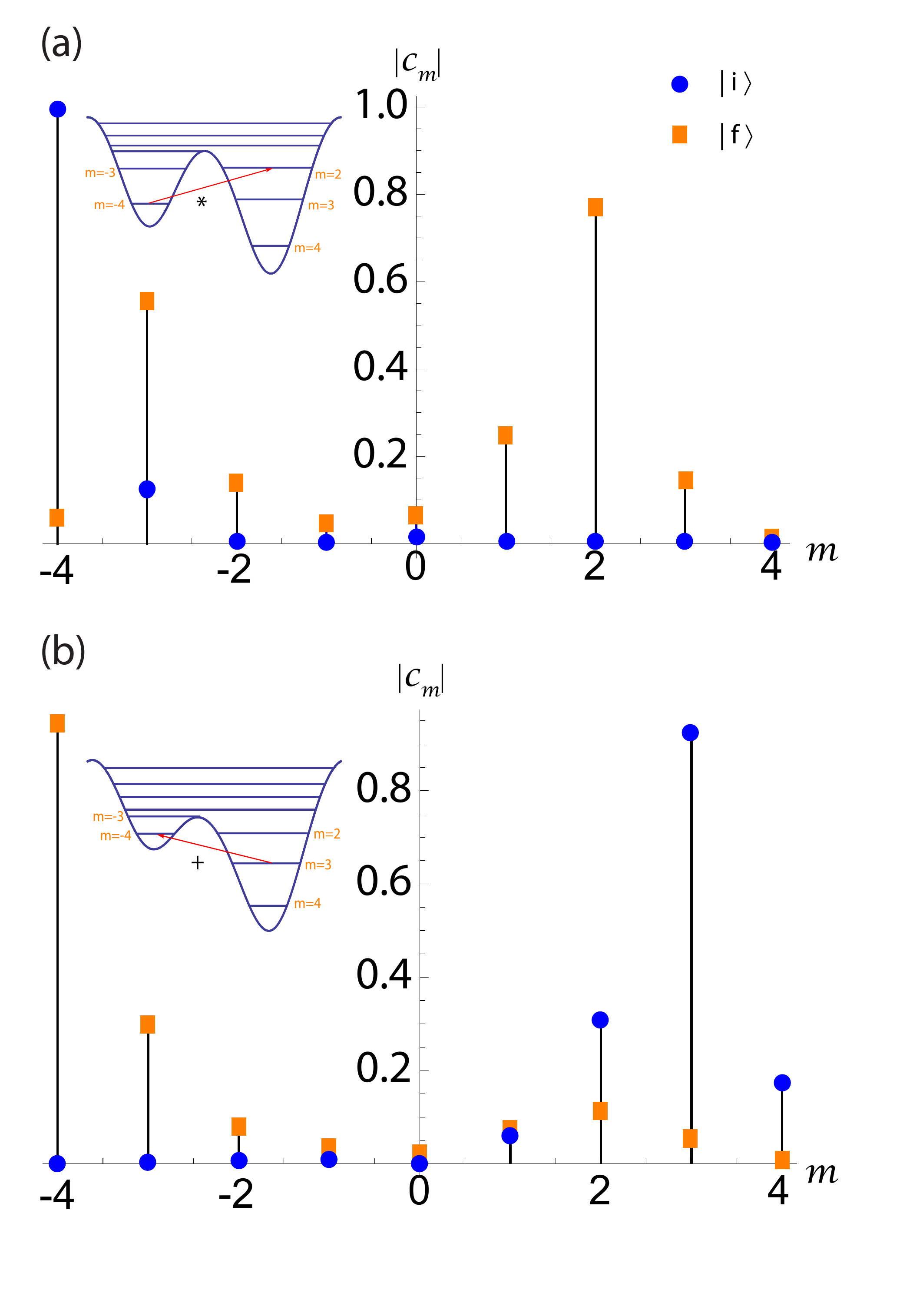}
\caption{Decomposition of eigenstates involved in the forbidden transitions in the $m$ basis (cf.~Eq.(\ref{basis})). Values of $\left|c_m\right|$ were calculated by diagonalizing the spin Hamiltonian at the fields corresponding to the (a) $\star$ and (b) $+$ transitions, setting $\theta=30^\circ$.  Blue circles (orange squares) indicate the values of $\left|c_m\right|$ for the lower (upper) state involved in each transition.  Insets schematically show the double-well potentials for the associated transitions, marked with red arrows.   }\label{forbidden2}
\end{figure}

The forbidden transitions (orange arrows in Fig.~\ref{Energygraph}) are observable because each occurs at a field near an anticrossing, where resonant tunneling takes place. Tunneling effects can be demonstrated by expanding the two energy eigenstates for each forbidden transition in the eigenbasis of $S_z$:
\begin{equation}
\ket{E_j}=\sum_m{c_m^{(j)} \ket{m}}.
\label{basis}
\end{equation}
Figures \ref{forbidden2}(a) and \ref{forbidden2}(b) show  $\left|c_m\right|$ vs.~$m$ for the initial ($\ket{i}$) and final ($\ket{f}$) states involved in the $\star$ and $+$  transitions, respectively, at $\theta = 30^\circ$ in the proximity of an anticrossing. For $\star$, $\ket{i}\approx\ket{m=-4}$, while  $\ket{f}$ is a superposition of primarily $\ket{m=2}$, $\ket{m=-3}$, and $\ket{m=1}$. It is the proximity of the ESR transition to an anticrossing produces a non-negligible amplitude of $\ket{m=-3}$ in $\ket{f}$ and thus a $\Delta m = 1$ transition matrix element between $\ket{i}$ and $\ket{f}$. The transition between states largely localized in separate wells constitutes a tunneling-assisted forbidden transition.
Equivalently, the  transition can be viewed as photon-induced tunneling in which the system transits between wells while absorbing the photon without acquiring enough energy to surmount the barrier.
During this forbidden transition, the change of $m$ is nominally $6$; indeed, a rigorous calculation
yields a change in expectation value
$\left|\Delta\langle S_z \rangle\right|$ as high as 6, indicating a large change in the spin's angular momentum with the absorption of a single photon~\cite{supp}.

Similarly, the $+$ transition (Fig.~\ref{forbidden2}(b)), involves $\ket{i}\approx\ket{m=3}$ and $\ket{f}$, a superposition of mostly  $\ket{m =-4}$, $\ket{m = -3}$, and $\ket{m = 2}$ states. This transition's proximity to an anticrossing here gives rise to a finite amplitude of  $\ket{m = 2}$ in $\ket{f}$ and a dipole matrix element with $\ket{i}$. For this transition, we calculate a maximum $\left|\Delta\langle S_z \rangle\right|$ of $\sim$7
for experimentally relevant values of $B$ and $\theta$~\cite{supp}.

The forbidden-transition peaks tend to become stronger when very close to allowed transitions (Fig.~\ref{spectrum}), confirming the delocalization of $\ket{f}$ near the tunneling resonance field.  A comparison of the experimental and simulated spectral intensity (Fig.~3
in~\cite{supp}) shows good agreement, with the intensity growing near anticrossings or at large transverse fields,
where tunneling is enhanced.

The peak linewidths for forbidden transitions tend to be significantly smaller than for allowed transitions (Fig.~\ref{spectrum}).  The widths appear to roughly scale as $1/\Delta \langle S_z \rangle$. This suggests that these peaks are homogeneously broadened. Extracting $T_2$ from the widths, yields values $\approx$ 0.1 -- 1 ns (Fig.~4
in~\cite{supp}), comparable to those found previously for Ni$_4$~\cite{delbarco_quantum_2004}.  Larger $T_2$ values are needed for realistic quantum information processing.  Long $T_2$ times have been achieved in a variety of molecular magnets via dilution~\cite{ardavan_will_2007,schlegel_direct_2008,bader_room_2014} to reduce dipole couplings; indeed, Ni$_4$ can be diluted by cocrystallizing it with the diamagnetic analog Zn$_4$~\cite{collett_Ni4_2016}.  $T_2$ can also be enhanced by making use of ``clock transitions", i.e.~operating near an anticrossing, where $\partial f/\partial B=0$ and decohering field fluctuations can only affect energies quadratically~\cite{bollinger_laser-cooled-atomic_1985,vion_manipulating_2002,wolfowicz_atomic_2013,shiddiq_enhancing_2016}. Nevertheless, the short  $T_2$ we observe may be compensated by the high density of Ni$_4$ molecules in a crystal that can enhance the spin-photon coupling~\cite{eddins_collective_2014}.

Independent of issues of coherence, the observed transitions have a distinctly ``macroscopic" character, involving states with largely different values of $m$. Linear superpositions between such states are prototypical examples of macroscopic superposition states (\emph{\`{a} la} Schr\"{o}dinger's cat). Here we characterize the observed transitions
$\ket{i} \longrightarrow \ket{f}$ in terms
of the linear superposition
$\left|\psi\right\rangle = \left(\ket{i} + e^{i\eta}\ket{f}\right)/\sqrt{2}$ that can be generated through pulsed excitations, where $\eta$ typically depends on time.
The ``macroscopicity" of such states can be quantified using suitable measures, such as the quantum Fisher information (QFI)~\cite{troiani_size_2013}:

\begin{equation}
\mathcal{F}_\psi=\max_{X,\eta}{\left[\bra{\psi}X^2\ket{\psi}-\bra{\psi}X\ket{\psi}^2\right]},
\label{QFIeq}
\end{equation}
Up to a constant (which we omit), QFI equals the variance of the operator $X =
\sum^N_{i=1} \mathbf{n}_i \cdot \mathbf{s}_i$, where the $\mathbf{s}_i$ refers to the $i$th ionic spin of the molecule.  $\mathcal{F}_\psi$ is maximized over all  possible  unit vectors $\mathbf{n}_i$ and with respect to the phase $\eta$. Here we
consider states belonging to the maximal-spin multiplet
($S$ = 4) of the Ni$_4$ molecule. One can show that in
this case the maximum is always obtained with parallel
vectors ($\mathbf{n}_i  = \mathbf{n}, \forall i$).

\begin{figure}[h]
\centering
\includegraphics[width=1\linewidth]{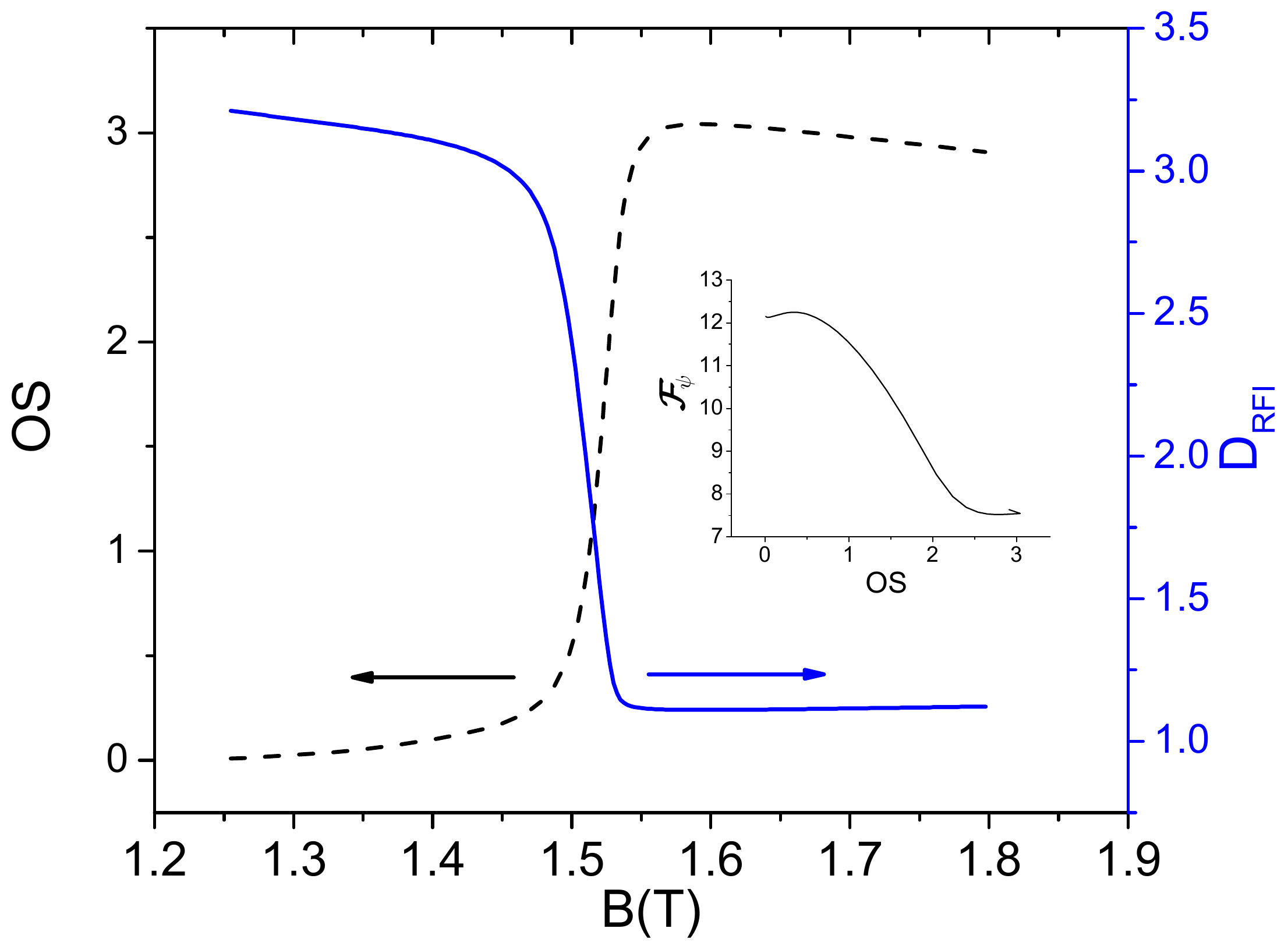}
\caption{Oscillator strength (OS) and $D_{RFI}$ for one of the transitions studied as a function of field.  Here $\ket{i}\approx\ket{m=3}$ and $\ket{f}=\ket{E_3}$ are the second- and third-lowest energy eigenstates (cf.~Fig.~\ref{Energygraph}), respectively.  As the field increases, the angle $\theta$ is adjusted to maintain resonance of the transition with the radiation frequency.  For this pair of levels, the transition is forbidden (allowed) at small (large) fields with a crossover at the field of the anticrossing.  The inset shows a parametric plot of $\mathcal{F}_\psi$ vs.~OS, illustrating how, near the anticrossing, one quantity rises as the other falls, but both can be substantial over some region.  Similar calculations for the transition between $\ket{E_2}$ and $\ket{E_4}$, the second and fourth energy eigenstates, show complementary behavior \cite{supp}.
}
\label{QFI}
\end{figure}

We also determine the relative Fisher information:
\begin{equation}
D_{RFI}=\frac{\mathcal{F}_\psi}{\frac{1}{2}\left[\mathcal{F}_i+\mathcal{F}_f\right]}
\end{equation}
in which each $\mathcal{F}$ is maximized
independently. The above normalization allows one to single out the amount of quantum fluctuations in $\ket{\psi}$ that result from the linear superposition of the states $\ket{i}$ and $\ket{f}$.
Figure \ref{QFI} shows calculated oscillator strength (OS, transition matrix element squared) and $D_{RFI}$ for the $+$ transition of the black component between $\ket{i}=\ket{E_2}\approx\ket{m=3}$ and $\ket{f}=\ket{E_3}$, the second and third lowest energy eigenstate, respectively, as a function of field.  $\theta$ is adjusted to maintain the resonance condition between the radiation frequency and the transition, following the right dashed black curve in Fig.~\ref{BvsTheta}.  At large fields, $\ket{f}\approx\ket{m=2}$, the transition between these levels is allowed with a large OS and $D_{RFI}\approx1$.  At low fields, $\ket{f}\approx\ket{m=-4}$ and the transition is more macroscopic ($D_{RFI}\approx3$) and forbidden (OS small).  Near the anticrossing, where states with very different values of $m$ hybridize, relatively
large values of $D_{RFI}$ can persist, while the oscillator strength remains finite.  Interestingly, the behavior of $D_{RFI}$ and $\left|\Delta\langle S_z \rangle\right|$ are qualitatively similar~\cite{supp}, indicating that $\Delta\langle S_z \rangle$ is a reasonable proxy for quantifying the macroscopicity of these transitions.

Our work demonstrates the important role tunneling can play in ``opening up" forbidden transitions.  In Ni$_4$, the relevant tunnel splittings for the transitions studied are relatively large (on the order of 1~GHz).  As a consequence, $m$ is no longer a good quantum number near an anticrossing, enabling forbidden transitions with large $\left|\Delta\langle S_z \rangle\right|$ and macroscopicity.  In addition, the large tunnel splittings allow tunneling effects to extend beyond the immediate vicinity of an anticrossing.  In our experiments, the observed forbidden transitions lie slightly away from anticrossings, permitting \emph{direct} single-photon transitions between states largely localized in opposite wells.  When tunnel splittings are much smaller, one enters the regime of photon-assisted tunneling~\cite{sorace_photon-assisted_2003,bal_radiation-_2008}, where an allowed ESR transition is followed sequentially by tunneling between wells.  Tunnel splittings can be enhanced by applying large transverse fields~\cite{delbarco_quantum_2004}. However, a field only acts as a perturbation when the Zeeman energy is small compared to molecule's anisotropy energy. In the large-field regime, the transitions become allowed and the macroscopicity of superposition states becomes suppressed.  Furthermore, going beyond the perturbation regime undermines the advantages afforded by clock transitions.  The tunnel splittings found intrinsically in Ni$_4$ are sufficient to observe forbidden transitions without the need of applying significant transverse fields to enhance tunneling.

\begin{acknowledgments}
We are indebted M.~P.~Sarachik for useful discussions and  comments on the manuscript.  We also thank D.~I.~Schuster and M.~Foss-Feig for productive conversations.  We are grateful to H.~Xu for her assistance in some aspects of the numerical simulations.  We thank C.~Euvrard (Millitech Corp.) for technical help with some equipment, J.~Kubasek for assistance in fabrication of the resonator and N.~Page for technical support of measurement equipment.  Support for this work was provided by the U.~S.~National Science Foundation under Grant No.~DMR-1310135, by the Italian Ministry of Education and Research through the FIRB project RBFR12RPD1, and by the EU through the FP7 FET project MoQuaS (contract N.610449). J.R.F.~acknowledges the support of the Amherst College Senior Sabbatical Fellowship Program, funded in part by the H.~Axel Schupf '57 Fund for Intellectual Life.  R.A.A.C.~thanks CNPq for the fellowship that enabled his work at the University of Massachusetts Amherst.
\end{acknowledgments}

\bibliography{transition_8_20_2015}

\clearpage
\newpage


\section*{Supplementary material}

# Supplementary Material for "Observation of tunneling-assisted highly forbidden single-photon transitions in a Ni$_4$ single-molecule magnet"

Yiming Chen,[1,2] Mohammad D. Ashkezari,[1] Charles A. Collett,[1] Rafael A. Allão Cassaro,[3] Filippo Troiani,[4] Paul M. Lahti,[3] and Jonathan R. Friedman[1]

[1]*Department of Physics and Astronomy, Amherst College, Amherst, MA 01002-5000, USA*
[2]*Department of Physics, University of Massachusetts, Amherst, MA 01003, USA*
[3]*Department of Chemistry, University of Massachusetts, Amherst, MA 01003, USA*
[4]*S3 Instituto Nanoscienze-CNR, I-41124 Modena, Italy*
(Dated: September 29, 2016)

## EXPERIMENTAL DETAILS

### Synthesis and Sample Preparation

Samples of Ni$_4$ were produced using published procedures [25]. Single crystals of typical dimension $\sim 1$ mm were produced and crystal structure verified through unit cell parameters using X-ray diffractometry.

### Measurement apparatus and procedure

We performed reflection ESR spectroscopy using a custom-built experimental probe within a Quantum Design PPMS. An Agilent 83650B signal generator was used as microwave source. A combination of a 4x active multiplier and a passive doubler was employed to cover the operational frequency range of 100-150 GHz. The resulting microwave signal was injected into a resonant cavity via a $\sim$1-m-long WR-10 waveguide (gold-plated stainless steel) with $\sim$3 dB insertion loss). The cavity is located in the sample chamber of the apparatus at cryogenic temperatures.

A custom-built 3D cylindrical resonant cavity was machined out of oxygen-free copper (cavity dimensions: diameter = 3.38 mm, depth = 6.86 mm). ESR measurements utilized the cavity's TE$_{011}$ resonant mode with resonant frequency of $\sim$115.54 GHz at low temperature and a quality factor ($Q$) of $\sim$10,000. A small quartz rod was placed along the axis of the cavity to break the degeneracy of the TE$_{011}$ and TM$_{111}$ modes. The cavity was coupled to the end of the waveguide through a 0.76-mm-thick copper coupling plate with a center coupling hole of diameter 0.64 mm. A static magnetic field $\vec{H}$ was applied along the axis of the cavity. A single crystal of Ni$_4$ was mounted onto the bottom of the cavity with grease. The rf field at the sample position was along the radius of the cavity, perpendicular to the static field. The easy axis of the crystal was manually tilted at various angles ($\theta_H$) relative to the static field, $\vec{H}$. The precise values of $\theta_H$ were determined from the spectra through fitting, as described below. Data was collected from three different samples. The spectrum labeled 15.2° in Fig. 2 (main text) was taken using Sample 1; the spectra labeled 26.6 − 38.6° were taken from Sample 2; the remaining spectra come from Sample 3. The angle $\phi_H$ (angle between hard (x) axis and transverse (x-y)

component of $\vec{H}$) was hard to control in mounting the sample. Tilting the sample to change $\theta_H$ had minimal effect on $\phi_H$ for a given sample so $\phi_{H,i}$ was taken to be constant for each sample $i$.

We measured the reflected power as a function of frequency, recording the dependence with an oscilloscope. After subtraction of a smooth frequency-dependent background signal (primarily due to waveguide resonances), a Lorentzian fit was applied to extract the resonance frequency and $Q$ value of the cavity at each field, resulting in one data point on the dispersive spectrum (see below) and $Q$-value spectrum (Fig. 2, main text), respectively.

### Inductive Offset Correction

When obtaining our spectra, the magnetic field $H$ is swept at a constant rate of 150 Oe/s. Since the sample chamber and cavity are metallic (mostly copper), the changing field produces an induced magnetic field that opposes $dH/dt$ (Lenz' law). Therefore, the actual $H$-field applied to the sample is less than the nominal value determined by the current in the magnet coils. To characterize this offset, we took some spectra by sweeping $H$ from -20 kOe to 20 kOe. An example is shown in Supplementary Fig. 1. One can see that the spectrum is not symmetric about $H = 0$, indicating the effect of the induced field. By determining the symmetry point for this spectrum, we found the inductive field offset to be $H_{off} = 301$ Oe. The spectra in Fig. 2 (main text) have been corrected from the raw data by shifting the spectra horizontally (to the left) by $H_{off}$.

## ADDITIONAL RESULTS

### Dispersive Spectra

We acquired ESR data by monitoring the cavity-sample resonance as the magnetic field was swept. From the resonance peak (absorbed power as a function of frequency $f$) at each value of field, we extracted values for $Q$ and $f_{res}$, the system's resonant frequency. The absorption spectra in the main text show $Q$ as a function of $H$. Dispersive spectra of $f_{res}$ as a function of $H$ allow us to extract similar information about

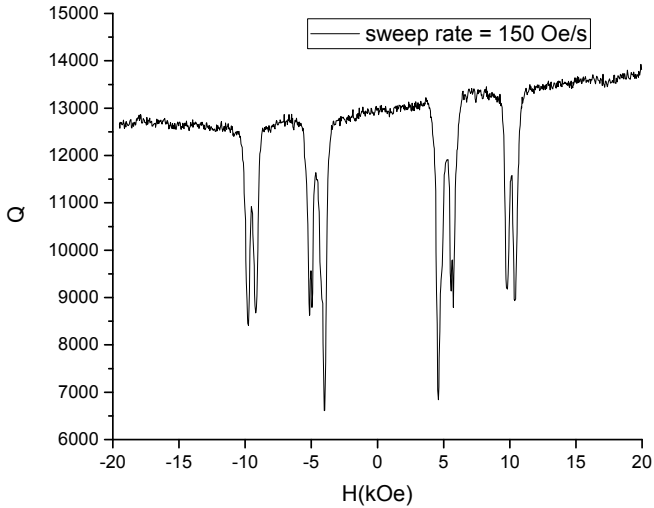

Supplementary Figure 1. ESR spectrum for $\theta_H = 15.2°$ with $H$ swept from -20 kOe to 20 kOe.

sample-cavity resonance fields. An example of one of these spectra is shown in Supplementary Fig. 2. As expected, a sample-cavity resonance in this spectrum is characterized by an up-down dispersive response in $f_{res}$, centered on the resonance field. Each peak in an absorbtion spectrum corresponds to a response in the associated dispersive spectrum. In this example, one can discern five ESR transitions (in order of increasing field): two allowed transitions, followed by a small forbidden transition (marked), then two more allowed transitions.

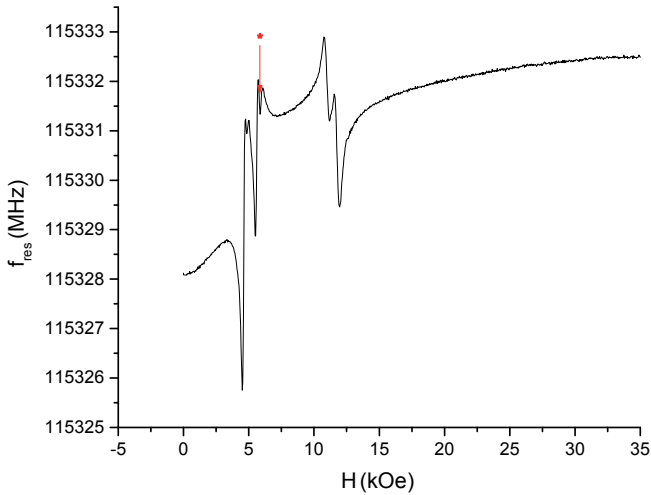

Supplementary Figure 2. Dispersive ESR spectrum for $\theta_H = 26.6°$. The mark indicates a forbidden transition.

One interesting feature that can be discerned only in the dispersive spectrum is the sign of the $f_{res}$ dependence on field: most ESR transitions show $f_{res}$ to increase initially as the field is increased and then to decrease rapidly when the field passes through the sample-cavity resonance field (anomalous

dispersion). This is a result of the Zeeman effect causing levels to move apart as the field is increased. The resonances in Supplementary Fig. 2 at ~13 kOe show this typical behavior. These transitions correspond to the right black arrow in Fig. 1 (main text), where indeed the two levels involved in the transition are moving apart as field increases, i.e. these are levels in the right well in the inset to Fig. 4 (main text). In contrast, the resonances in Supplementary Fig. 2 at ~5 kOe have the opposite dispersive behavior, with $f_{res}$ first decreasing with $H$ and then rapidly increasing when resonance is reached. This unusual behavior indicates that the two levels involved in these transitions are moving *closer* to each other as field increases. The left black arrow in Fig. 1 (main text) shows that this transition corresponds to levels exhibiting just that behavior, as they are levels in the left well of Fig. 4 (inset, main text).

## Spectral Intensity

We determine the spectral intensity of each peak in each spectrum by measuring the total area of the peak after subtracting background. We compared this with the calculated transition intensity $|\langle f| S_T |i\rangle|^2$, where $S_T = \vec{S} \cdot \hat{B}_{rf}$. We normalized each quantity so the total intensity for all transitions in a given spectrum is unity. Supplementary Figures 3(a) and 3(b) show a comparison of the experimental (points) and simulated (curves) spectral intensity of the $\star$ and $+$ forbidden transitions, respectively, as a function of $\theta$. Supplementary Figure 3(a) contains the $\star$ transition intensity for the "black" component only since the corresponding transition for the "red" component typically overlaps with the allowed transition for the black component or is just too small to be discerned, depending on $\theta_H$. For the $+$ transition, the peaks from the two components overlap with each other (cf. Fig. 3, main text) and cannot be easily distinguished. So, Fig. 3(b) shows the combined intensity for both components. Both panels show good agreement (excepting for $\theta > 50°$ in Fig. 3(b)).

## Peak Widths

Inhomogeneous broadening due to a distribution of dipolar fields would produce peaks of nearly the same width. In contrast, we find that the forbidden peaks are significantly narrower than the allowed ones, suggesting homogeneous broadening. Supplementary Figure 4 shows values of the decoherence time $T_2$ extracted from the measured linewidths as a function of $\theta$ for the four transitions (two allowed and two forbidden) examined in this study. To obtain $T_2$, we fit the peaks in each spectra in Fig. 2 (main text) to Lorentzians to determine the width $\Delta H$ (in Oersted). After converting to a width $\Delta B$ (in Gauss), we used the calculated field dependence of the transition frequency, $\partial f/\partial B$ to determine $T_2 = (\frac{\partial f}{\partial B}\frac{\partial B}{\partial H}\Delta H)^{-1}$. The analysis was done for the peaks associated with the "black" component (cf. Fig. 3, main text). The data are generally consistent with $T_2 \approx 0.1 - 1$ ns. While

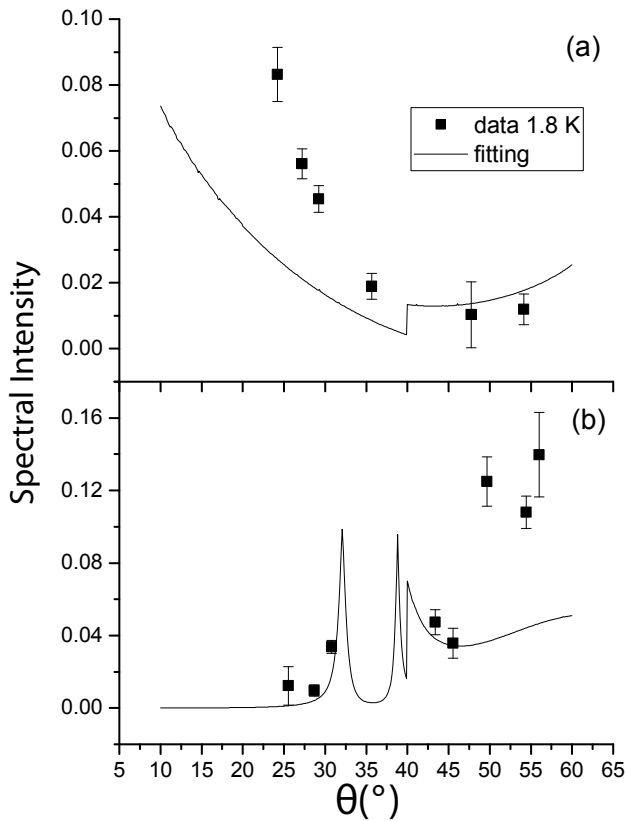

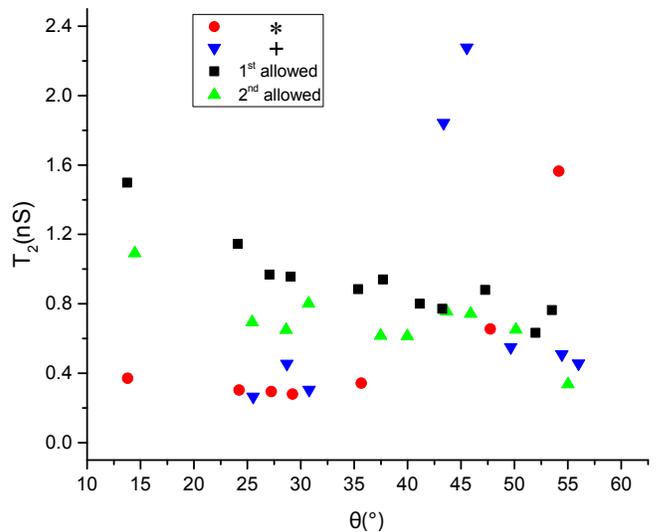

Supplementary Figure 4. Decoherence time $T_2$ for allowed and forbidden transitions, as indicated, as a function of $\theta$. Data was extracted from peak widths of spectra in Fig. 2 (main text). Most of the data presented is for the "black" component only. For the + transition, the peaks for the two components overlap at most values of $\theta_H$ and cannot be distinguished. Under those circumstances, $T_2$ values for that transition represent the width of the combined peak for both components.

Supplementary Figure 3. Spectral intensity as a function of $\theta$ for the (a) ★ and (b) + forbidden transitions. Points are experimental data determined from the area of the associated peaks. Curves are simulation results based on calculated transition matrix elements. For (a), the data and simulations are for the "black" transitions only. In contrast, because the peaks for + transitions substantially overlap, the experimental and calculated data in (b) are the combined intensity for both components. Error bars indicate standard errors obtained from fitting. The jumps in the calculated curves at $\theta = 40°$ arise from the fact that the value of $\phi_H$ suddenly changed when switching from Sample 2 to Sample 3.

there appears to be some systematic dependence of $T_2$ on $\theta$ and differences depending on the particular transition, it is hard to draw strong conclusions from these observations since the results depend on calculated values of $\frac{\partial f}{\partial B}$, which are sensitive to the values of Hamiltonian parameters used. Small changes in these parameters would result in non-negligible differences in the values of $T_2$ that we can infer.

### Forbiddenness and Macroscopicity

A basic way to characterize the forbidden transitions observed in this study is to calculate the change in the expectation value of $S_z$ when a photon takes the system from state $|i\rangle$ to state $|f\rangle$. For an allowed transition, $\Delta\langle S_z\rangle \simeq \pm 1$ while for forbidden transitions, this quantity is expected to be significantly larger. Supplementary Fig. 5 shows calculated values of $|\Delta\langle S_z\rangle|$ as a function of $\theta$ for the four transitions (two allowed and two forbidden) studied using parameters for the black component. The value of $B$ is adjusted as $\theta$ is varied so that the resonance condition with the applied radiation frequency is maintained for each transition. For each transition in the figure, the value of $|\Delta\langle S_z\rangle|$ changes from $\sim 1$ to a much larger value, up to $\sim 7$. This change occurs as the final energy eigenstate of the transition passes through an anticrossing and the character of the transition switches from being allowed to forbidden or *vice versa*. The calculations were done for two values of the azimuthal angle $\phi_H$, corresponding to the values for Samples 2 and 3. The arrows in the figure indicate experimental conditions for which forbidden transitions were observed (Fig. 2, main text), showing that some of these transitions have very large values of $|\Delta\langle S_z\rangle|$.

A more rigorous way to quantify the macroscopicity of superposition states is the quantum Fisher information, $\mathcal{F}$, as discussed in the main text. We find that superposition states that are actualizable through exciting the forbidden transitions, can have a reasonably large relative Fisher information, $D_{RFI}$. This is shown in Fig. 5 (main text) for the transition between energy eigenstates $|E_2\rangle$ and $|E_3\rangle$, where the latter state changes character as the changing field $B$ (and $\theta$) pass the state through an anticrossing and the transition changes from forbidden to allowed. Complementary behavior is seen for the transition between states $|E_2\rangle$ and $|E_4\rangle$, as shown in Supplementary Fig. 6. In this figure, we again see a correlation between small oscillator strength (OS) and large $D_{RFI}$, the hallmarks of a forbidden transition.

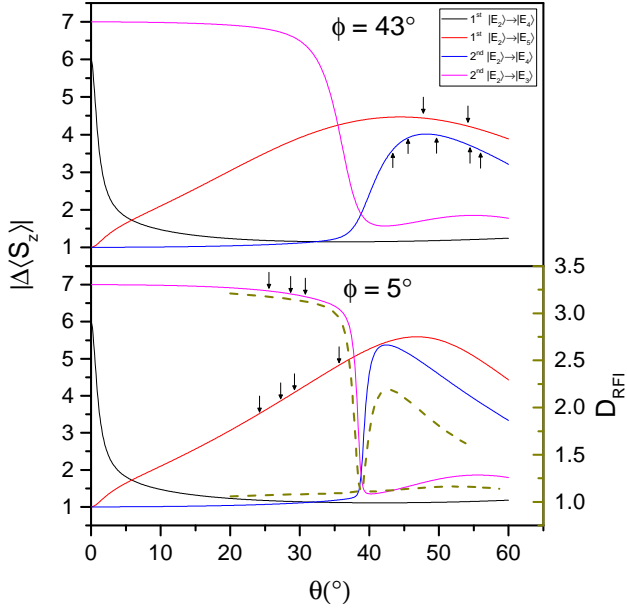

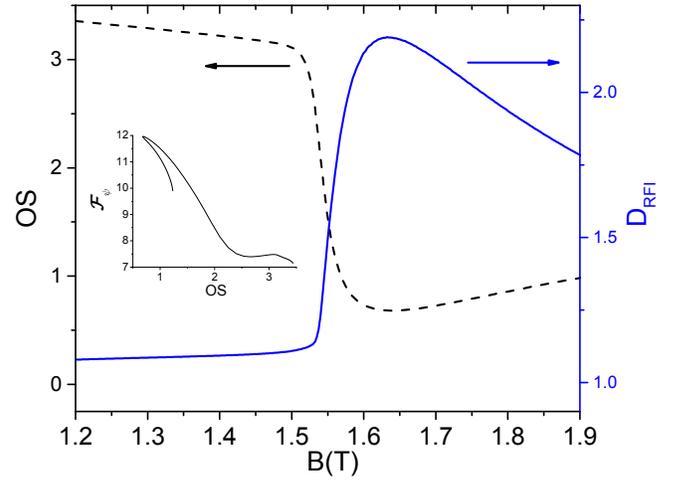

Supplementary Figure 5. Calculated $|\Delta\langle S_z\rangle|$ as a function of $\theta$ for the transitions studied. Calculations were performed for $\phi_H = 5°$ and $43°$, as indicated, corresponding to the orientations of Samples 2 and 3, respectively. The labels "$1^{st}$" and "$2^{nd}$" refer to the transitions that occur at lower and higher fields, respectively (cf. Fig. 3, main text). Note that for the $1^{st}$ transitions, $|E_2\rangle \simeq |m = -4\rangle$ while for the $2^{nd}$ transitions, $|E_2\rangle \simeq |m = 3\rangle$. Arrows indicate experimental conditions where forbidden transitions were observed. For comparison, the dashed lines give calculated values of $D_{RFI}$ for superpositions that can be created using the states involved in the $2^{nd}$ allowed and forbidden transitions.

We compare the two measures $|\Delta\langle S_z\rangle|$ and $D_{RFI}$ in the lower panel of Supplementary Fig. 5, where $D_{RFI}$ for the $2^{nd}$ allowed and forbidden transitions are shown by the dashed lines. There is a clear qualitative similarity between the behavior of $|\Delta\langle S_z\rangle|$ and $D_{RFI}$ although they are not quantitatively related by any simple transformation.

## DETAILS OF DATA ANALYSIS AND FITTING PROCEDURES

### Calculation of $\vec{B}$

Because the ESR spectra were obtained at low applied field and at substantial angles $\theta_H$ between the sample's easy (z) axis and $\vec{H}$, the field experienced by a typical molecule, $\vec{B}$, is not collinear with $\vec{H}$. Thus, it was important to carefully transform $\vec{H}$ into $\vec{B}$ in performing our analysis and simulations.

While for any given spectrum $\theta_H$ is constant as the field is swept, the angle $\theta$ between $\vec{B}$ and the easy axis is changing as $H$ increases because of the spin's anisotropy. To account for this, we diagonalized the Hamiltonian with $\vec{B}$ along the

Supplementary Figure 6. Calculated oscillator strength (OS) and $D_{RFI}$ for the transition between $|E_2\rangle$ and $|E_4\rangle$, the second and fourth lowest energy eigenstates, respectively, as a function of $B$. The angle $\theta$ is adjusted to maintain resonance of the transition with the radiation frequency. The inset shows a parametric plot of $\mathcal{F}_\psi$ vs. OS. Calculations were done using the parameters for the black component and using $\phi_H = 5°$, corresponding to Sample 2. The hairpin turn in the left of the inset arises from OS being a nonmonotonic function of $B$.

x, y, and z directions. For each direction, we calculated the magnetization $M_i$ ($i = x, y, z$) as a function of $B_i$ and temperature $T$ using standard statistical mechanical techniques. (The calculated $M_i$ were nearly identical for the black and red components.) The molecule's symmetry implies that the susceptibility tensor is diagonal for $H=0$; this becomes only an approximation as $H$ increases. We then used the relations

$$H_i = B_i - 4\pi\alpha M_i\left(B_i, T\right),\qquad(1)$$

where $\alpha$ is a parameter on the order of unity that takes into account lattice-structure and crystal-shape (demagnetization) effects and is treated as a free parameter in our fitting (see below). For a given $\theta_H$ and $\phi_H$, we calculate $H_i$ using:

$$\begin{aligned}
H_z &= H\cos\theta_H \\
H_x &= H\sin\theta_H\cos\phi_H \qquad(2)\\
H_y &= H\sin\theta_H\sin\phi_H,
\end{aligned}$$

Using Eq. 2 and numerically inverting Eq. 1, we can calculate $B_i\left(H, \theta_H, \phi_H, T\right)$ and thereby calculate $B = \left|\vec{B}\right|$ and $\theta = \arccos\left(B_z/B\right)$, yielding the values plotted in Fig. 3 (main text). We could also find the angle $\phi = \arctan\left(B_y/B_x\right)$. However, the small hard-plane anisotropy of Ni$_4$ means that the susceptibility is nearly isotropic within the plane, resulting in $\phi$ being nearly indistinguishable from $\phi_H$.

In simulating the spectra, $B$ and $\theta$ vary as a function of $H$ for fixed $\theta_H$. Through the fitting procedure described below, we determined best-fit values of $\alpha$, $\theta_H$ and $\phi_H$. We treated $\alpha$ as a parameter that is independent of sample orientation.

The spectrum labels in Fig. 2 (main text) are the values of $\theta_H$ determined from optimization.

## Fitting of Spectral Peaks

Spectra could readily be simulated using the EasySpin package [35]. However, fitting of the experimental spectra are complicated by the fact that the forbidden transitions are typically very small compared with allowed transitions. Therefore, a least-squares minimization procedure that compares experimental and simulated spectra effectively ignores forbidden-transition peaks for most values of $\theta_H$. Instead, we adopted a different fitting procedure that treats all observed peaks with equal weight. By fitting the experimental spectral peaks to Lorentzian functions, we determined the field position, $B_{data,i}$, and area, $I_{data,i}$, for the $i$th peak. A least-squares fitting method was then applied to determine anisotropy parameters $D$ and $A$ for both components as well as the g tensor, $\alpha$, and the angles $\theta_H$ and $\phi_H$. The anisotropy parameter $C$ was independently determined from low-frequency ($\sim$5 GHz) ESR measurements of the zero-field tunnel splitting of $\sim$ ($|m = 2\rangle + |m = -2\rangle$) and $\sim$ ($|m = 2\rangle - |m = -2\rangle$) states for each component (data to be presented elsewhere [29]). The $\chi^2$ for each peak is defined as:

$$\chi_i^2 \left( \theta_H, \phi_H \right) = \left( \frac{B_{sim,i} \left( \theta_H, \phi_H \right) - B_{data,i}}{\Delta B_{data,i}} \right)^2 \\ + \left( \frac{I_{sim,i} \left( \theta_H, \phi_H \right) - I_{data,i}}{\Delta I_{data,i}} \right)^2, \quad (3)$$

where $\Delta B_{data,i}$ and $\Delta I_{data,i}$ are the uncertainties in $B_{data,i}$ and $I_{data,i}$, respectively, as determined from the peak fitting. $B_{sim,i} \left( \theta_H, \phi_H \right)$ and $I_{sim,i} \left( \theta_H, \phi_H \right)$ are the values determined from simulations of the corresponding peaks for given values of $D$, $A$, $\mathbf{g}$, $\alpha$, $\theta_H$ and $\phi_H$. For the + forbidden transitions, the peaks from the two components overlapped for many values of $\theta_H$. So, in calculating $\chi^2$ for those peaks, we included the mean position and the total area of the peaks.

The total $\chi^2$ was calculated by summing over all the peaks from all spectra:

$$\chi_{total}^2 = \sum_i \sum_{\theta_H, \phi_H} \chi_i^2 \left( \theta_H, \phi_H \right) \quad (4)$$

$\chi_{total}^2$ was minimized to give the optimized values of anisotropy parameters and $\alpha$ as well as all values of $\theta_H$ and $\phi_H$. Anisotropy parameters and $\alpha$ were taken to be the same for all peaks from all spectra (with one exception – see below). Peaks from the same spectra were forced to have the same value of $\theta_H$. Spectra from the same sample were forced to have the same value of $\phi_H$. (Unlike for the other samples, the value of $\phi_H$ for Sample 1 was found to be a poor fitting parameter since the value of $\theta_H$ for the lone spectrum from this sample is small. So, $\phi_H$ for this spectrum was simply fixed at a value that corresponds to a weak minimum of $\chi^2$ for this parameter and the fit was performed varying the other parameters.) The resulting fitting parameters were used to correct the calculated values of $M_i$, as described above. The above procedure was then iterated until it converged.

$\chi_{total}^2$ was calculated by summing over all the peaks from all spectra at 115.54 GHz at temperatures 1.8 K and 9 K, as well as spectra (not shown) taken at $\sim$5 GHz (using a different apparatus) both for a sample of $Ni_4$ and a dilute sample consisting of 5% $Ni_4$ cocrystallized with 95% diamagnetic $Zn_4$ [29]. For the spectrum from the dilute sample, $\alpha$ was allowed to take on a different value from the spectra for the undiluted samples. Our fit yielded the values of anisotropy parameters and g factors given in the main text and a value of $\alpha = 0.31(1)$ for the undiluted samples. In Supplementary Figs. 7–17, we compare our experimental and simulated spectra at 1.8 and 9 K for $f = 115.54$ GHz. Each figure indicates the values of $\theta_H$ and $\phi_H$ found from the fitting procedure. The comparisons show that the simulations reproduce the experimental data well. Since our fitting procedure involves only peak positions and areas, the peak width was manually adjusted in these simulations (parameter HStrain in EasySpin) to achieve reasonable agreement. As can be seen in the figures, the width at 9 K is roughly a factor of three larger than at 1.8 K. This increase in width with temperature may correspond to a reduced lifetime of the excited states due to higher acoustic-phonon populations at higher temperature. We note that the broad feature at $B \sim 3.8$ T (observed in spectra that probed fields that high) is due to impurities in the apparatus and unrelated to the sample, as can be readily seen by the fact that the feature does not change as $\theta_H$ changes or when switching between samples.

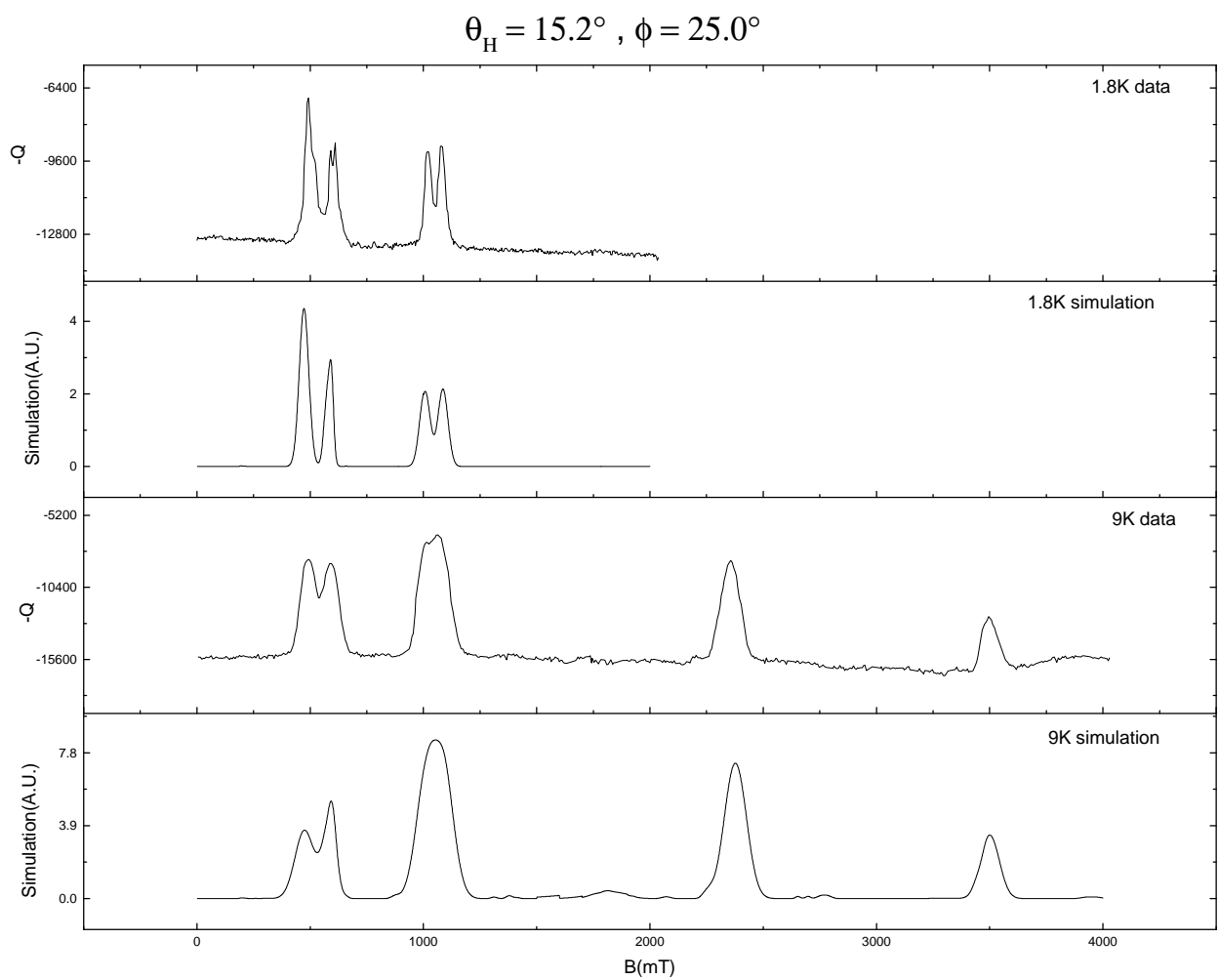

Supplementary Figure 7. Comparison of experimental and simulated spectra at 1.8 and 9 K, as indicated for $\theta_H = 15.2°$.

$\theta_H = 26.6°$ , $\phi = 5°$

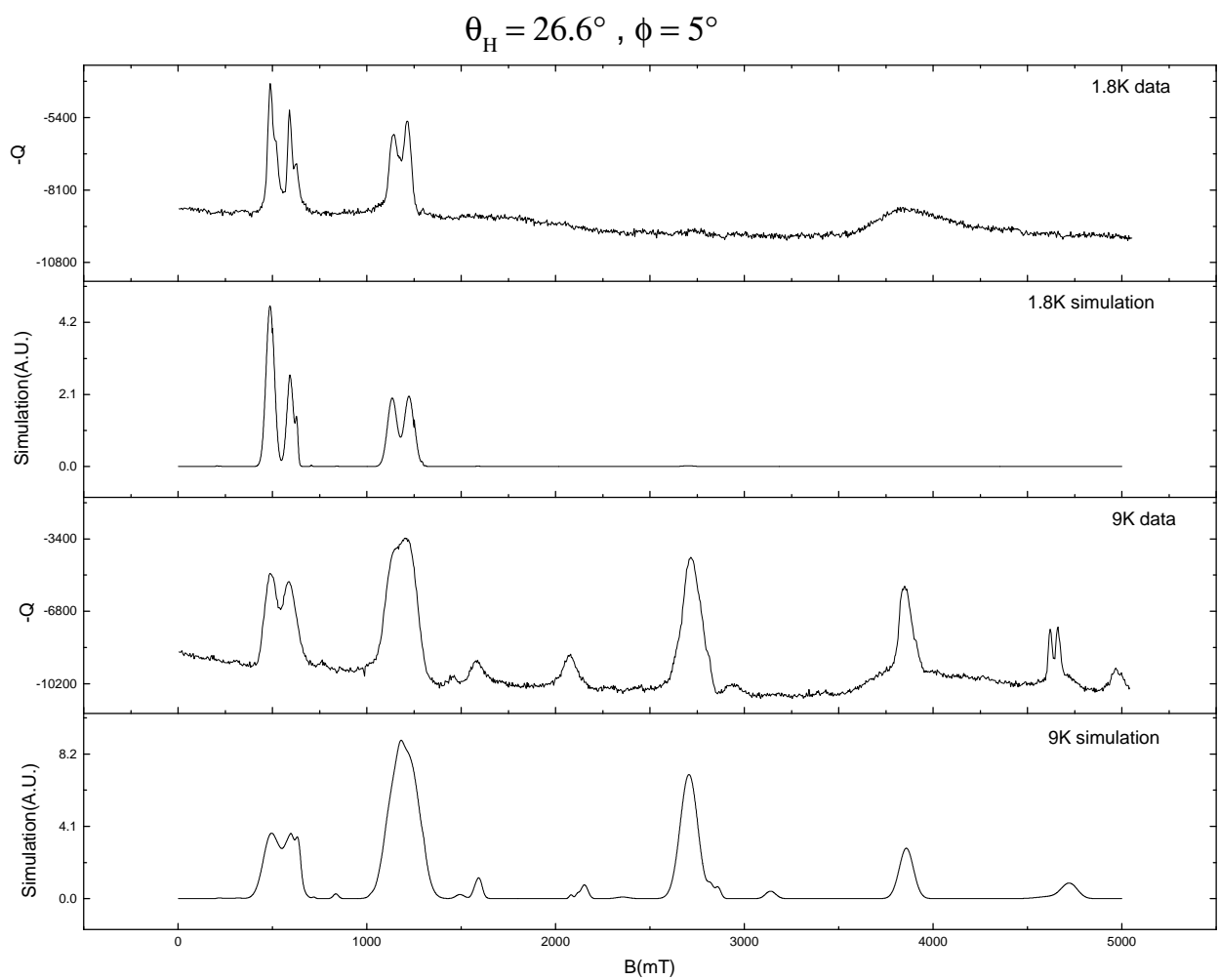

Supplementary Figure 8. Comparison of experimental and simulated spectra at 1.8 and 9 K, as indicated for $\theta_H = 26.6°$.

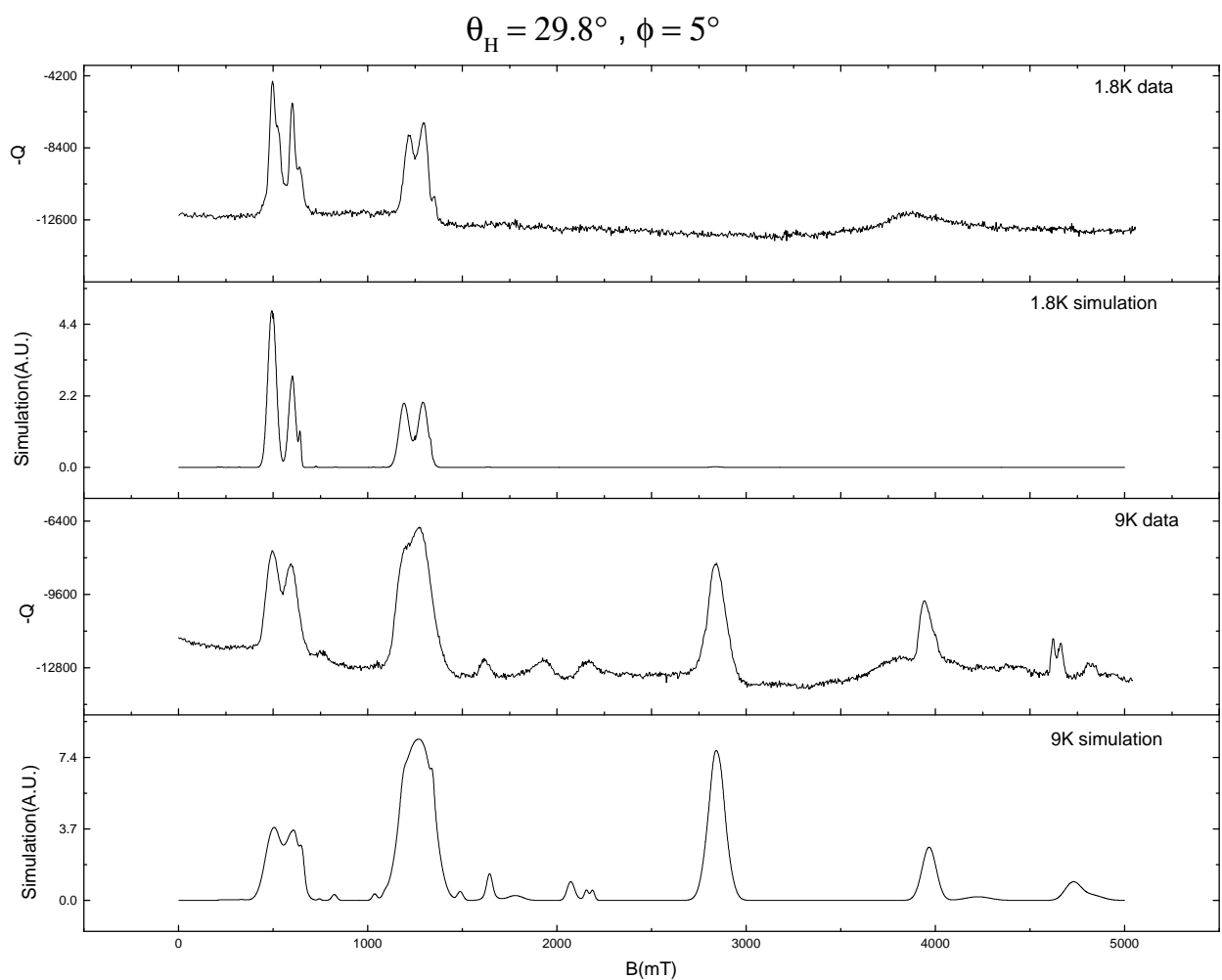

Supplementary Figure 9. Comparison of experimental and simulated spectra at 1.8 and 9 K, as indicated for $\theta_H = 29.8°$. Note that this spectrum was left out of Fig. 2 (main text) for clarity of presentation, but these data were included in the fitting and analysis.

$\theta_H = 32.0° \, , \, \phi = 5°$

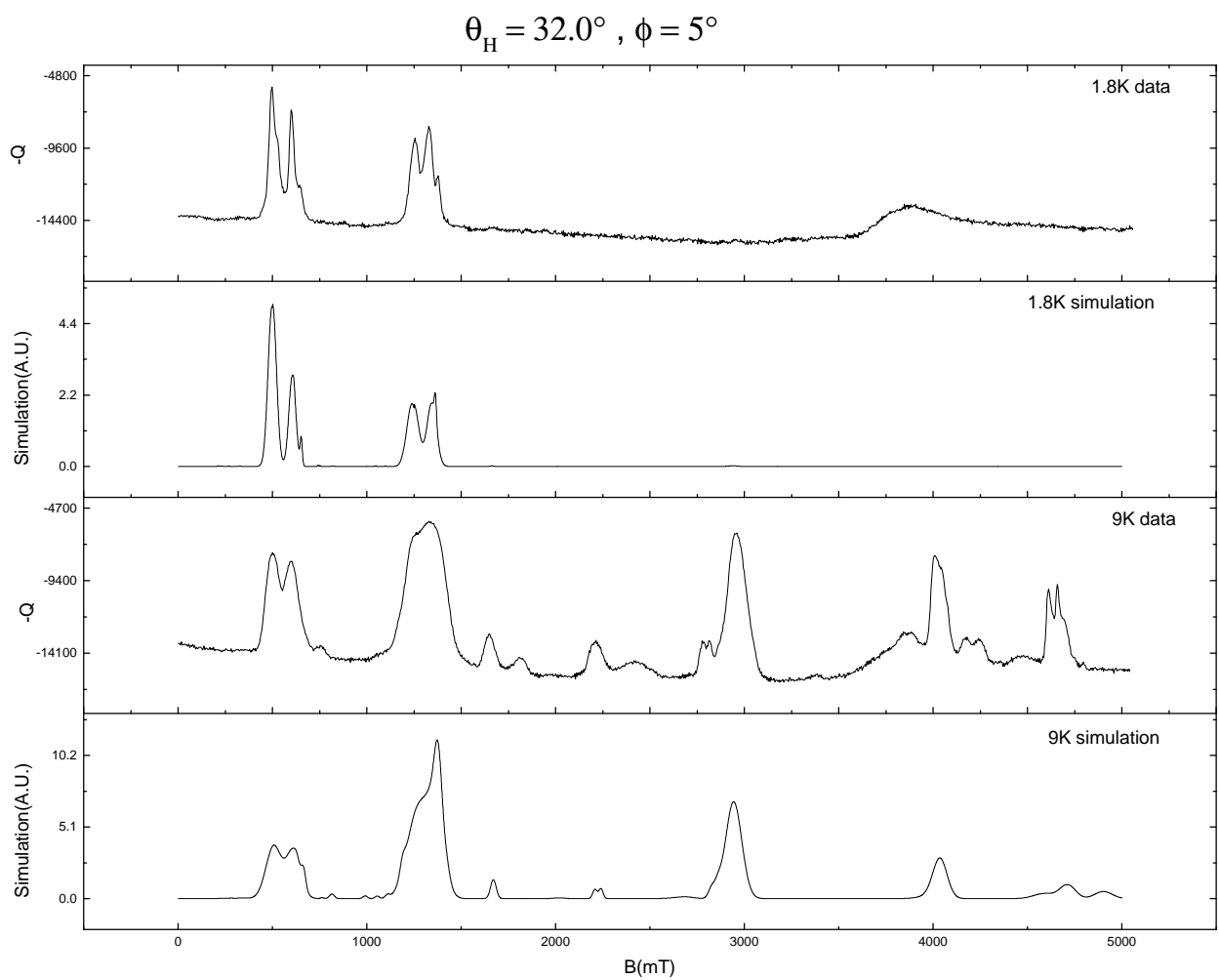

Supplementary Figure 10. Comparison of experimental and simulated spectra at 1.8 and 9 K, as indicated for $\theta_H = 32.0°$.

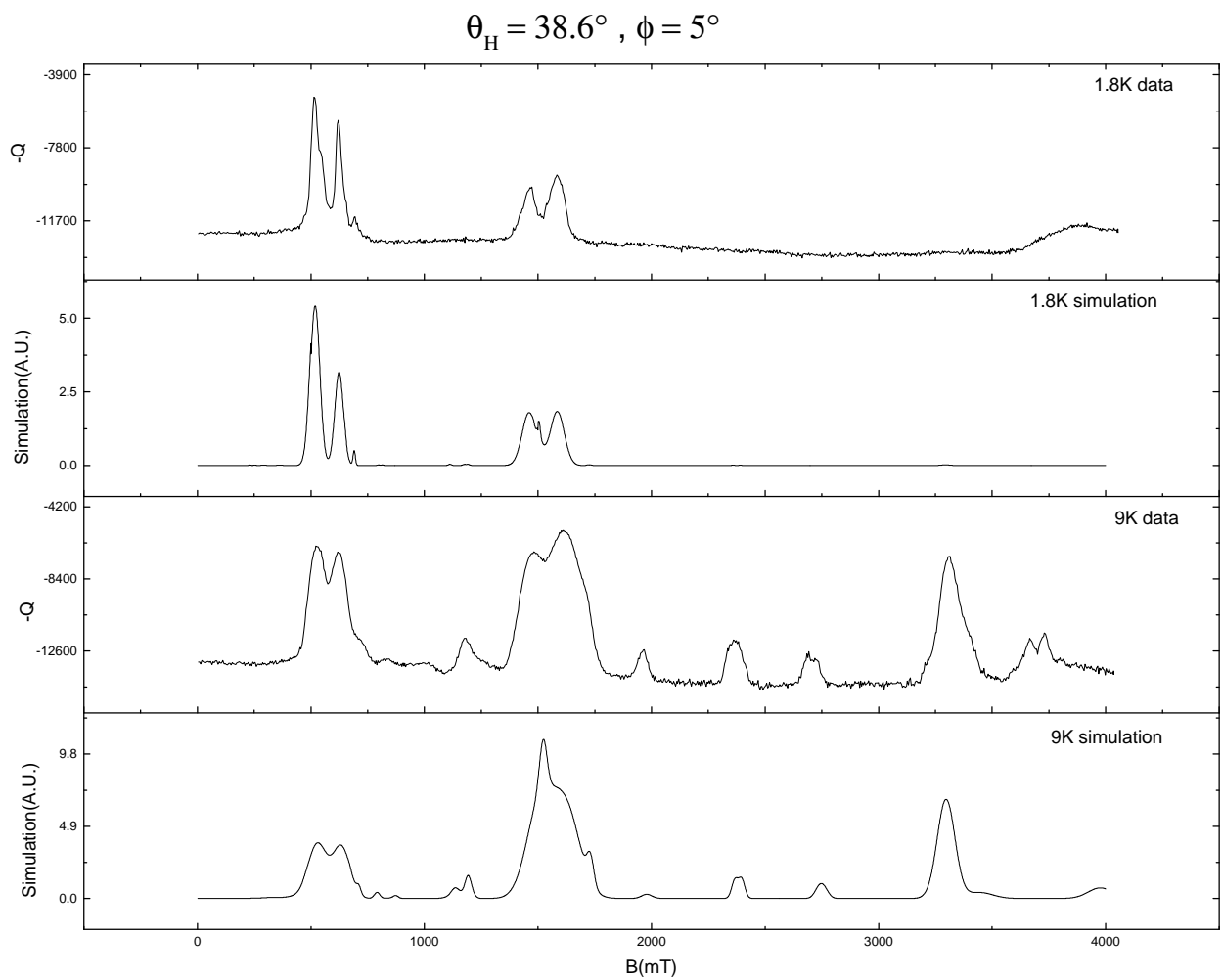

$\theta_H = 38.6° , \phi = 5°$

Supplementary Figure 11. Comparison of experimental and simulated spectra at 1.8 and 9 K, as indicated for $\theta_H = 38.6°$.

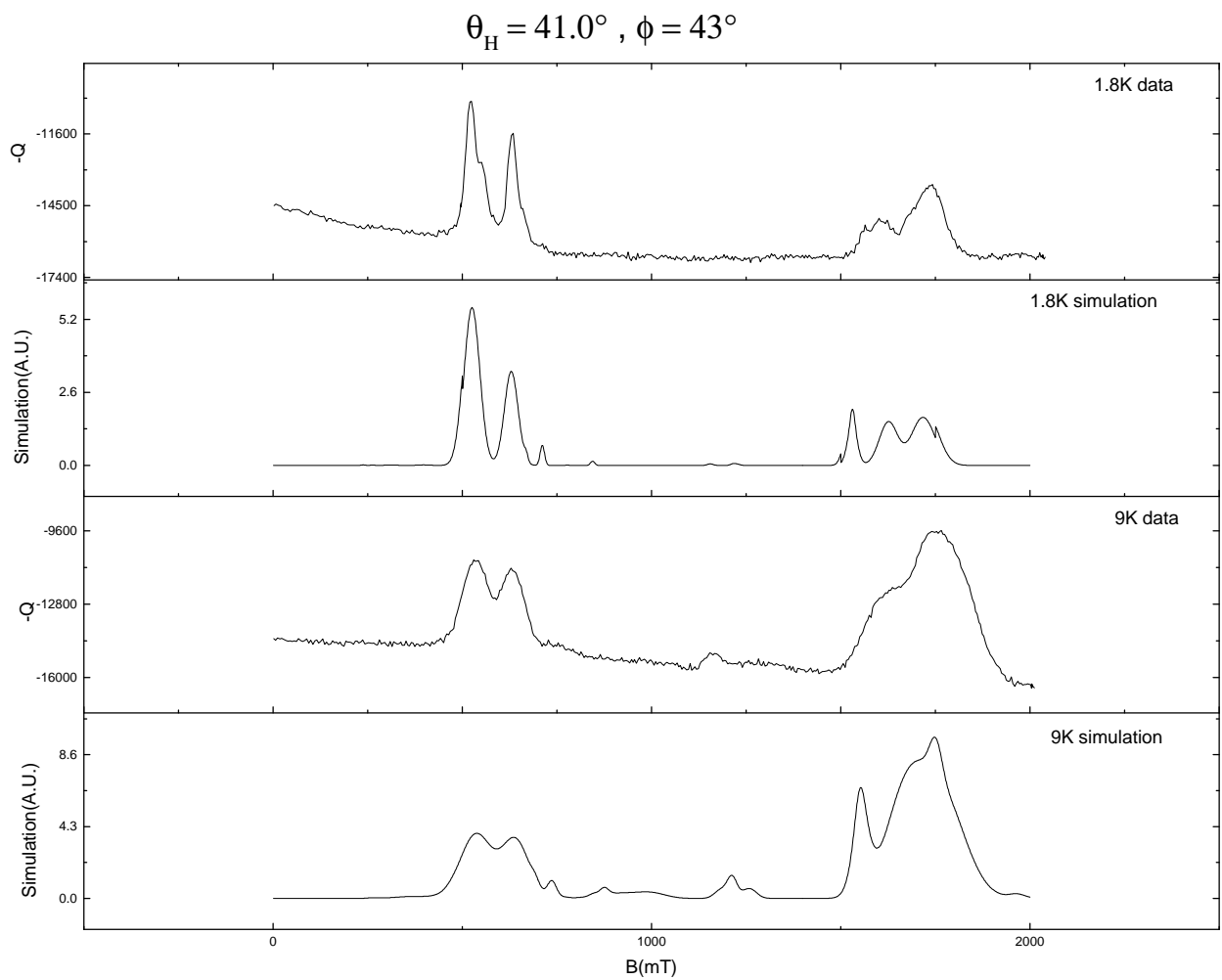

Supplementary Figure 12. Comparison of experimental and simulated spectra at 1.8 and 9 K, as indicated for $\theta_H = 41.0°$.

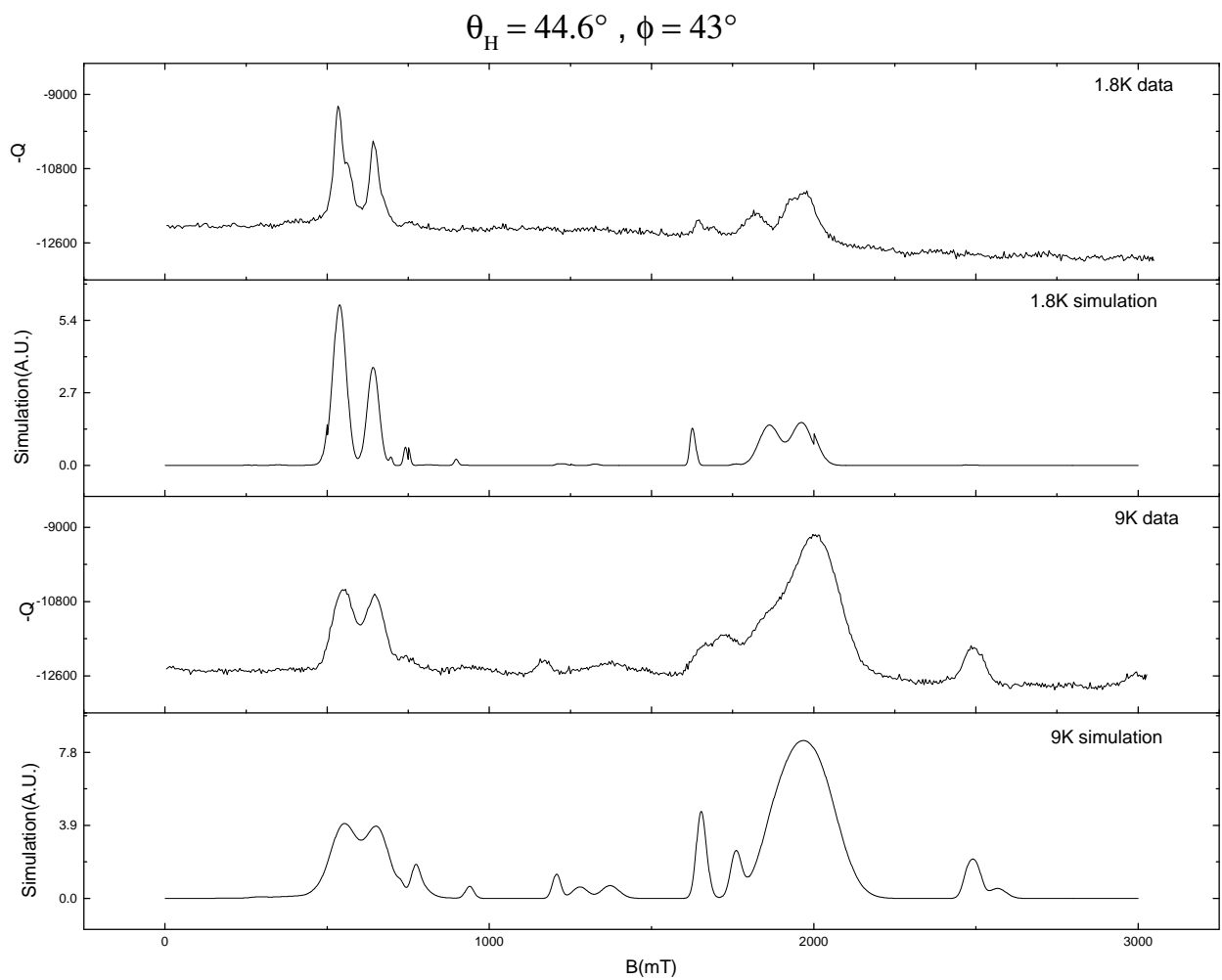

Supplementary Figure 13. Comparison of experimental and simulated spectra at 1.8 and 9 K, as indicated for $\theta_H = 44.6°$.

$\theta_H = 47.0°, \phi = 43°$

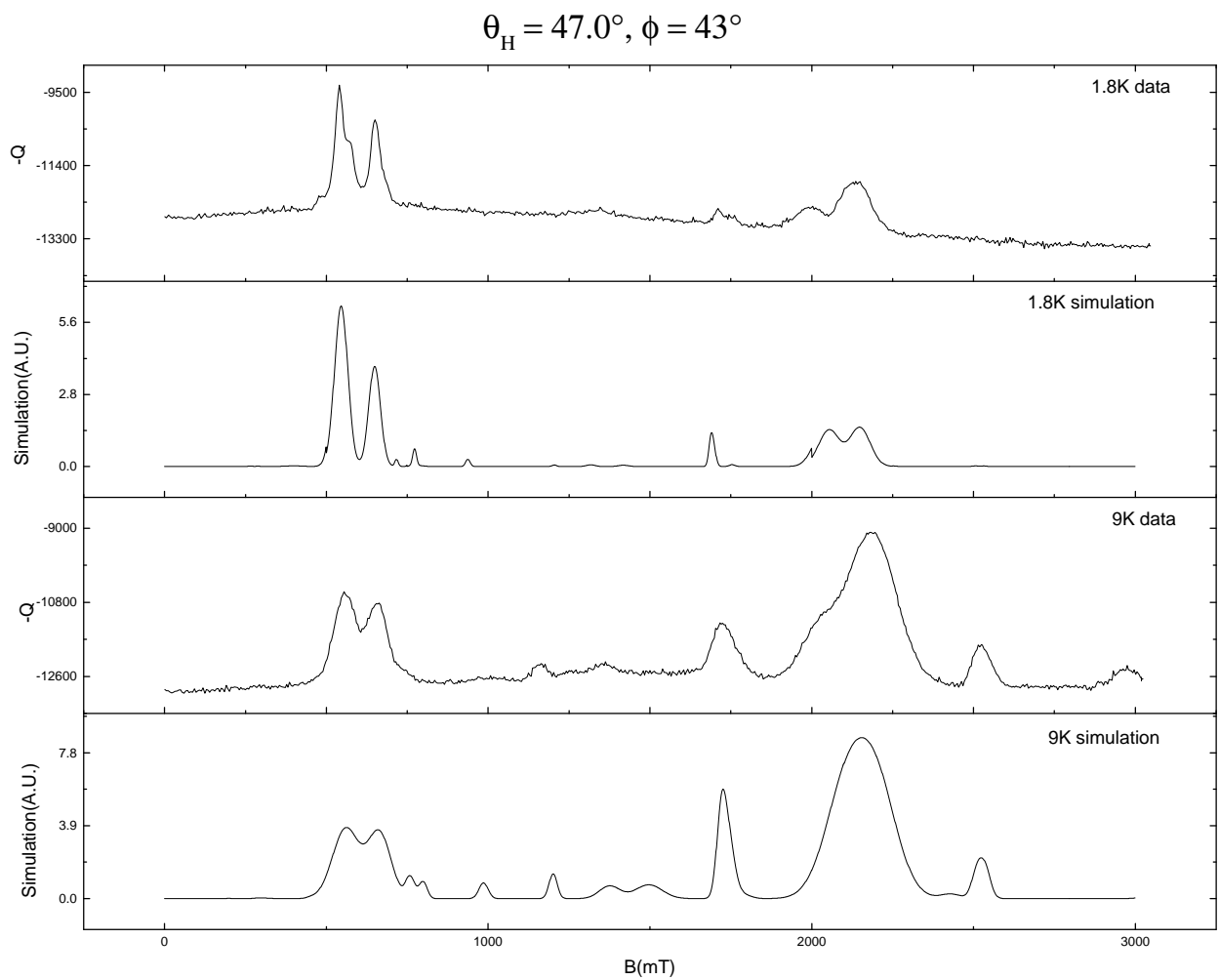

Supplementary Figure 14. Comparison of experimental and simulated spectra at 1.8 and 9 K, as indicated for $\theta_H = 47.0°$.

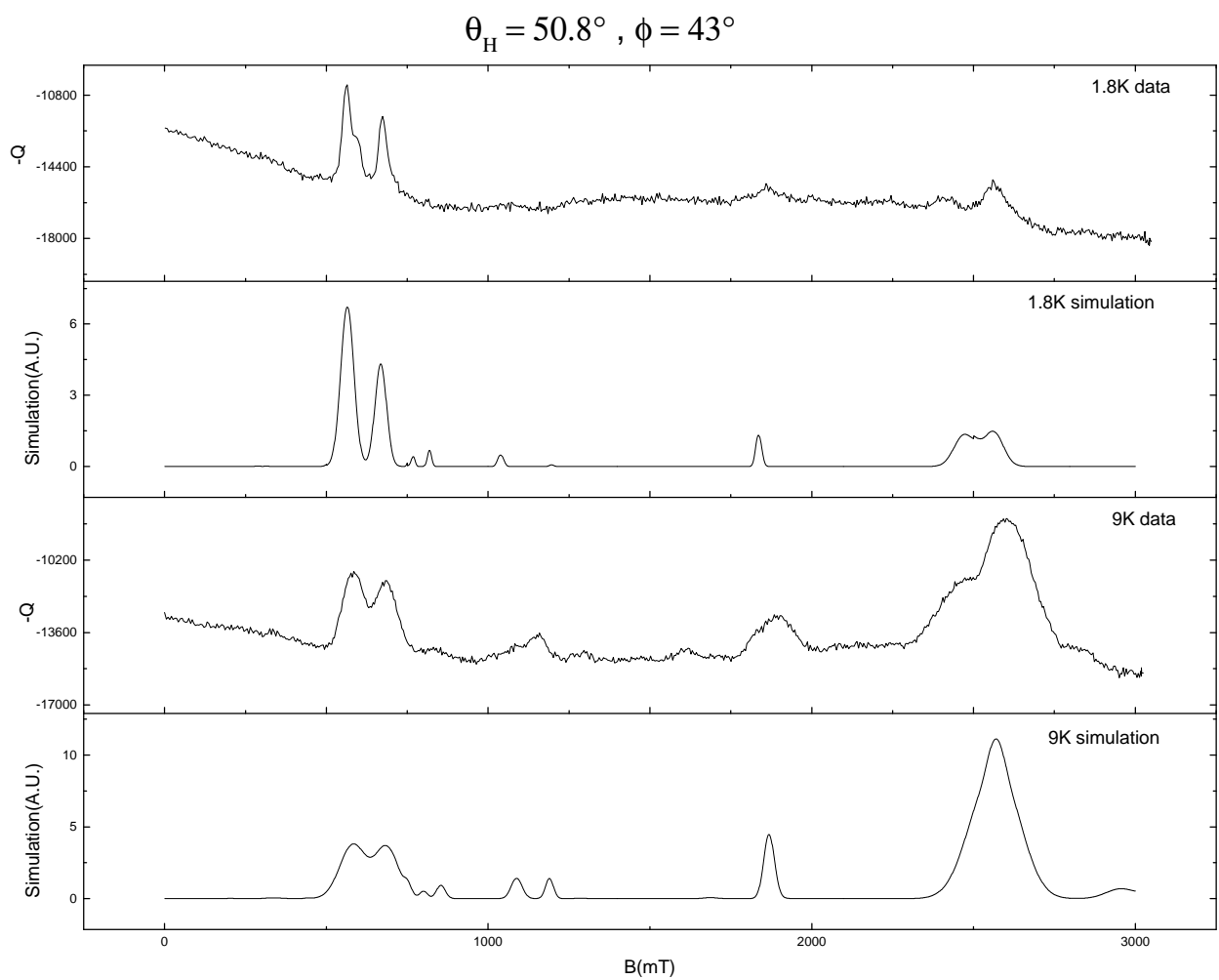

Supplementary Figure 15. Comparison of experimental and simulated spectra at 1.8 and 9 K, as indicated for $\theta_H = 50.8°$.

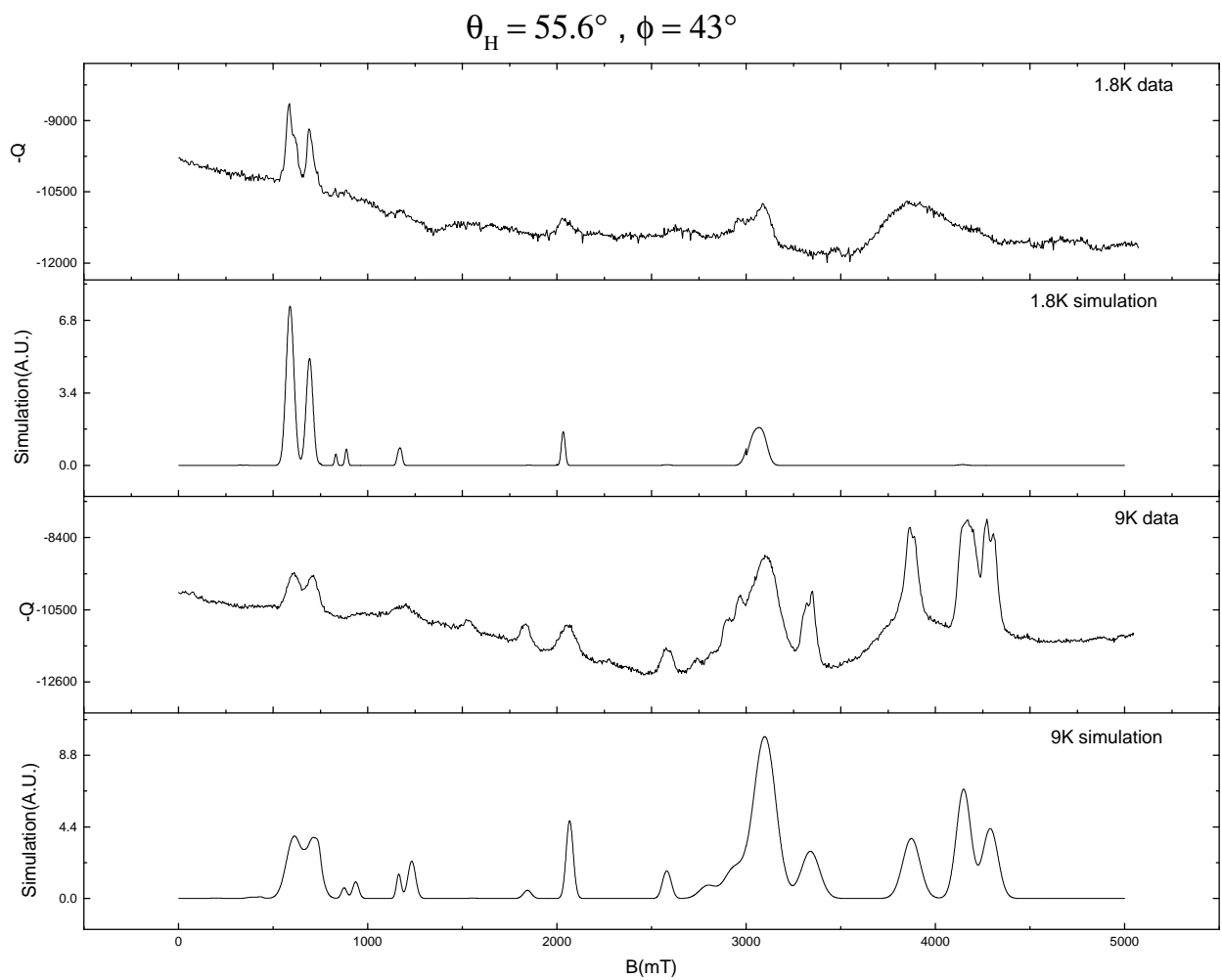

Supplementary Figure 16. Comparison of experimental and simulated spectra at 1.8 and 9 K, as indicated for $\theta_H = 55.6°$.

$\theta_H = 57.2° \, , \, \phi = 43°$

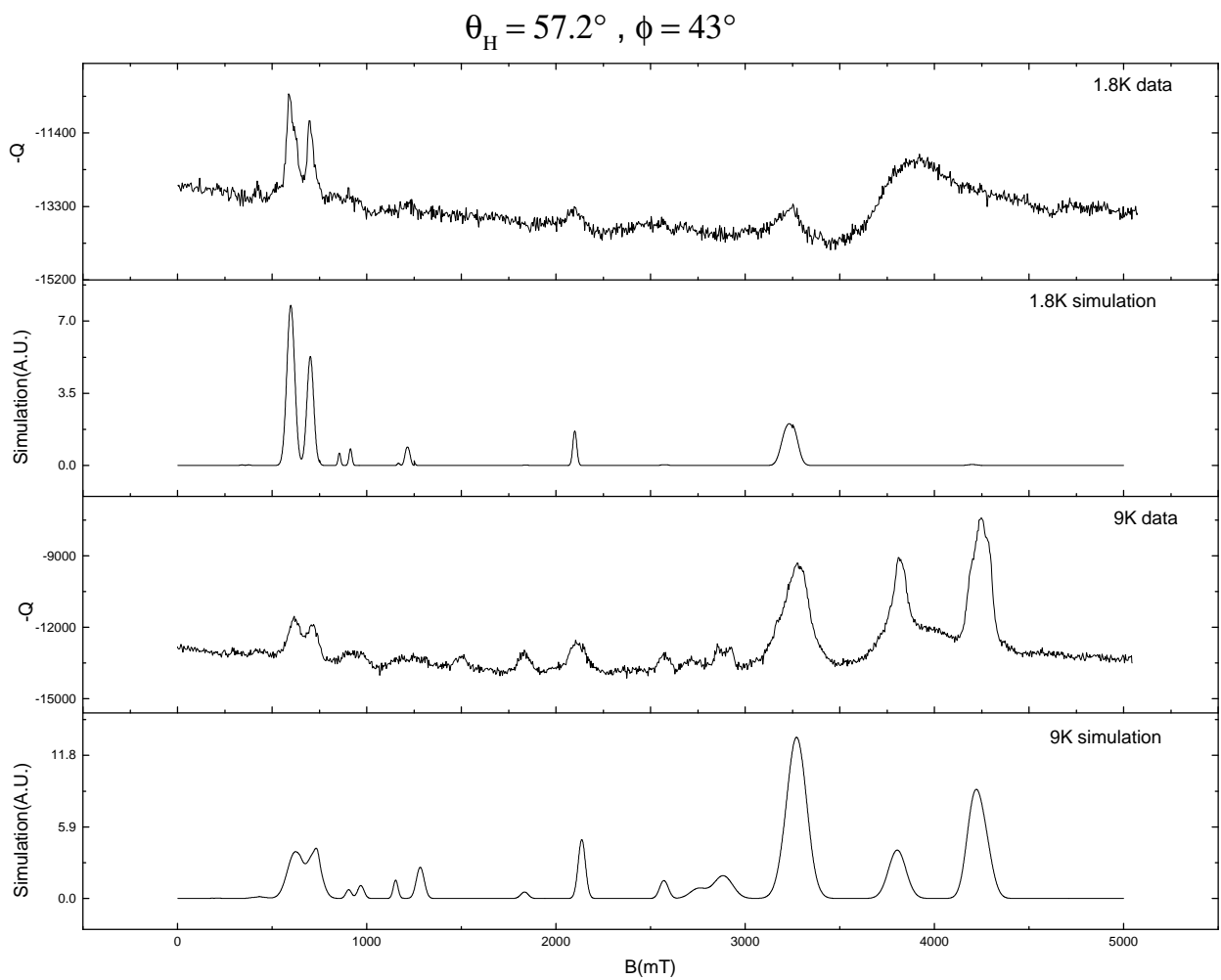

Supplementary Figure 17. Comparison of experimental and simulated spectra at 1.8 and 9 K, as indicated for $\theta_H = 57.2°$.



\end{document}